# Coupling of acoustic and drift modes, harmonic modons in astrophysical dusty plasma and a new mode in the Comet Halley

Zahida Ehsan[1,2], Nazia Batool[2] and Volodymyr M. Lashkin[3]

[1]*SPAR and The Landau-Feynman Laboratory for Theoretical Physics,*
*Department of Physics, CUI, Lahore Campus 54000, Pakistan.*

[2]*National Centre for Physics, Shahdara Valley Road, Islamabad 45320, Pakistan. and*

[3]*Institute for Nuclear Research, Kiev 03680, Ukraine*

(Dated: February 17, 2022)

# Abstract

This manuscript, presents a theoretical study of the linear and nonlinear characteristics of acoustic and drift waves in a bounded inhomogenious dusty plasma which has potential applications in space and lab environments. In this analysis, flow of all plasma particles is assumed to be along the axial direction whereas gradients in velocity are considered to be along the radial direction. First we study coupling of dust modified ion acoustic (DMA) and drift (DMD) waves (linear analysis) at fast time when dynamics of dust remain inactive, later formation of nonlinear vortex like structures are examined. In the second case at slow time when dust participates in dynamics, coupling between ultra low frequency dust acoustic (UDA) and dust drift (ULD) waves is studied and modons like solutions are obtained in the nonlinear analysis. Existence of the vortex solution with azimuthal harmonics higher than the dipole vortex has been studied both analytically and numerically. Later we extend our analysis to consider applications of dust particle's presence in the comet Halley and tropical mesospheric dusty plasma where in earlier case we report that that unexpectedly comet Halley plasma admits a new mode where presence of dust in background does not have any affect, this mode is similar to the convective cell mode where electrostatic drift ($E \times B$) cancels out, density gradients also vanish due to the presence of negative ions and only polarization drift contributes to the mode. For the linear analysis of this mode, there is no coupling observed but this the convective cell mode evolves whereas nonlinear dynamics of this mode are found to obey the vortex-like solution. In another example for the linear and nonlinear analysis, we model meteoritic dust in the mesospheric plasma where both primary and secondary electrons are taken into account where later are the source of positive dust charge shown by the observations. Formation of modons is found to depend upon the rotation and number density of the plasma particles. Moreover, we also extend our analysis for a special case of non-Maxwellian (Kappa and Cairns) dusty plasma, as space observations of particle velocity distributions indicate presence of either shoulders at low energy or suprathermal tail. Numerical analysis shows nonthermality significantly affecting the vortices. Significance of this analysis can be viewed fruitfully from the astrophysical and lab context.

## I. BACKGROUND

Modons (vortices) are long lived rotating structures in velocity pattern of a fluid which can be noticed in a broad range of the physical systems, ranging from quantum to galactic systems. They are frequently observed in the ocean, in atmosphere, in earth's ionosphere and magnetosphere. Vortices typically originate either by some external driving force e.g. by the spoon in the soup cup, planetary rotations whereas in plasma fluid, a combination of flow, localized heating in plasma experiments, electric current and magnetic field can be at the root of their formation. Modons accumulate energy and are often involved in the dynamic transport, whereas are also regarded as driving force in turbulence[1–4].

The most phenomenal example of localized planetary vortex is Great Red Spot on Jupiter. This vortex is significantly larger than earth in diameter. Vortices can be local or global and of ring shape as well, depending on the physical situation of the system. One important aspect of the interest in the study of vortical structure is that sometime growing vortex itself can be the possible free energy source and can accelerate plasma particles, one such example is formation of a ring vortex by the interaction of electromagnetic wave with dusty plasma eventually accelerating the dust particles to high velocity[5]. In addition some instabilities in plasma system may also lead to accumulation of energy at long scale length which can drive the formation of vortex like structures. Generally vortices are solutions of 2D nonlinear partial differential equations in the form of Bessel functions where Poisson bracket $[F, G] = \hat{z} \times \nabla F . \nabla G$ [where $F$ and $G$ are functions of field variables] is the source of nonlinearity.

Main motivation here is to investigate formation of modons in the dusty plasma for different environments this is because from more than three decades, dusty plasmas have proven to be an excellent research area for their tremendous applications in both space and laboratory technology (including fusion devices)[6–13]. Needless to mention experiments on dusty plasmas under microgravity conditions (which is now possible due to vehicles in space) provide an excellent opportunity for the better understanding of physical processes like phase transitions, collective excitations because of the large time scales associated with heavier big dust particles[14, 15].

Both positive and negative dust can naturally exist for instance in rings of planets, asteroid zones, comets, interstellar and circumstellar clouds, Earth's mesosphere and lower ionosphere etc. Composition of dusty plasma entirely depends upon the circumstances.

Our solar system contains billions of comets which are composed of frozen gases and cosmic dust particles revolve around the Sun. During revolution, when a comet reaches near Sun, it throws out dust particles and in this, ionization of gases increases thus this dust and gases together make a dusty plasma. This dusty plasma exists in the tail of a comet expands millions of miles outwards from the Sun[16–19]. Dust grains in comet become electrically charged by the plasma (such as solar wind) which permeates the solar system. Observations from Vega1 spacecraft revealed much of the carbon is trapped within dust grains. Comets in the solar system are believed to be the darkest objects because of their low geometric albedo ($\sim 0.04$). Comet Halley which can be seen after a gap of about 76 years is considered to be the most luminous comet. Observational data from Vega 2 spacecraft revealed Halley's surface has temperature of $300 - 400\ K$ and is shielded with a dust mantle. Trajectories of the micron-submicron dust grains which were to be released from the cometary nucleus have been studied. Observations from Vega 1 & 2 and Giotto spaceports revealed the presence of both positive and negative ions of hydroxides, hydrogen, oxygen, siliconetc. in the comae of comet Halley[20–23]. Thus presence of these ions other than dust and electrons make this *electronegative* plasma system.

Everyday a huge flux of interplanetary meteoroids with velocity $\sim 11.2 - 72.8\ (kms^{-1})$ enters into the ambience of planet Earth. During this, these meteoroids boil away in such a way to become part of the levitated dust particles in the mesosphere. Processes like photo- and thermionic emissions are responsible for making dust positively charged in mesospheric plasmas on the dayside. When observations were initiated to investigate charged dust in the tropical mesosphere, a layer of dust particles of 10 nm radii was detected near the mesopause. Observation taken half hour after local astronomical sunset showed an unexpected positively charged dust in the upper part of the layer whereas lower part contained usual negative dust. Reason behind this: dust was still carrying its pre-sunset positive charge due to photoemission. Later it was suggested that even on the night side, micrometeoroids in the upper mesospheric region before their ablating into the mesopause get positively charged due to thermionic electron emission[24–26].

At the same time, when solar wind passes through the astrophysical objects present in the solar system such as comets or the plasma in the vicinity of the planets, makes these enviornemnts (plasma system) more complex which nore follows usual state of equilibrium. In this situation, plasma species in particular electrons and ions will asymptotically lead to a nonequilibrium steady state for which power-law distributions need to be defined. For instance often Cairns [27], Kappa [28] or r, q distribution [29] are considered be best fitted.

New time and spacial scales arise when dust is present in the system of plasma which profoundly affects the behavior of linear and nonlinear processes[30–36].

Indeed dusty plasma in lab devices are of finite extent, also in principle the real plasma system contains inhomogeneities in density and so can drive drift motions associated to waves. Many problems in the past have been reported where bounded plasma systems were taken into account. Vranjes and Poedts made a rigorous analysis of waves propagating in the inhomogenous plasma and showed propagation is influenced by the geometry of the device where plasma system is under investigation[37]. This is to be noted that most of the known theories about dispersion of complex plasma systems should be valid only under particular circumstances.

In this paper we however will divide problem of acoustic and drift modes coupling (linear regime) and the subsequent formation of modons (nonlinear regime) for the two time scales in a dusty plasma. First when dust particles are static, stay in the background and do not participate in the dynamics however their presence will play a role via quasineutrality condition, we call this low frequency regime (fast time scale phenomena). Second is the case when dust will play an active role through its dynamics eventually arising ultra-low frequency phenomena at slow time scale. Drif and dust acoustic modes associated to the fast [and slow] time scales will be called as dust modified drift (DMD) [ultra low frequency drift (ULD)] and dust modified acoustic (DMA) [ultra low frequency dust acoustic (UDA)] wave. We will extend this analysis to the non- Maxwellian, comet Halley and tropical mesosphere plasma.

To the best of authors' knowledge this has not been studied earlier and results of the present investigation are useful for understanding the physics of formation of modons in several situations of dusty plasma.

The paper is organized in the following manner: In Sec. II, the Physical assumptions and description of the model is given. Section III deals with both the fast and slow time scale formulation of dust laden plasma. For both the cases, respective linear dispersion relations are obtained for the coupled dust modified drift and acoustic waves. Section IV deals with the nonlinear analysis and subsequent derivation of the Hasegawa- Mima equation and its solution. In Sec. V is devoted to study of Comet Halley Plasma however this only deals with a case when dust is present in the background. Section VI deals with the tropical mesospheric plasma. Section VII devotes to the formation of harmonic of modons for a non-Maxwellain (Kappa and Cairns) distributed dusty plasma whereas this restricts to only nonlinear analysis. Finally, conclusions are given in Sec. VI.

## II.  PHYSICAL ASSUMPTIONS AND DESCRIPTION OF THE MODEL

To formulate the physical problem of coupling of acoustic and drift wave and then formation of modons, we will make following the assumptions:

1. In our analysis we assume dust grains' size is much smaller than the mean distance $\left(r_d \sim n_d^{-1/3}\right)$ between the dust particles and spatial scales of the problem therefore this can be regarded as a three-component (dust laden plasma) system whose constituents are electrons $(n_e)$, ions $(n_i)$, and charged dust grains $(n_d)$. All dust particles are assumed to have same mass (and size) and charge; our premise is not that stringent though because within the fluid element variations in the mass and charge can happen. Also within a single system of dusty plasma, dust particles can have different sizes and shapes which impacts the charging process and associated physical processes like excitaion of waves and instabilities, overall it's too difficult to treat such a complicated system.

2. The plasma under consideration is magnetized, homogeneous, bounded and flowing along the external magnetic field $B_0 = B_0\hat{\mathbf{z}}$ in a cylinder of radius $r_0$. Dust particles are for simplicity assumed to have same mass and radius, and collisions between the particles are ignored. The quasineutrality condition is given as:

$$n_i(r) = n_e(r) \pm Z_d n_d(r) \qquad (1)$$

here $\pm Z_d$ is the charge of dust grain, depending upon the situation we will chose dust to be negative or positive. $n_s = n_{s0} + \delta n_s + \delta n_s^L$, while $n_{s0}$ represents the equilibrium density, the parameter $s$ denotes the species i.e., dust, ions and electrons. Superscript '$L$' refers to the ultra-low frequency for the dust acoustic wave in comparison (slow time scale) with the higher frequency dust modified acoustic (DMA) wave, and $\delta n_s$ gives the density perturbation on the fast time scale. The continuity and momentum equations are given respectively as

$$\frac{\partial n_s}{\partial t} + \nabla \cdot (n_s \mathbf{v_s}) = 0 \tag{2}$$

and

$$n_s \left(\frac{\partial}{\partial t} + \mathbf{v}_s \cdot \nabla\right) \mathbf{v}_s = \frac{q_s}{m_s} n_s \left(\mathbf{E} + \frac{\mathbf{V}_s}{c} \times \mathbf{B}_o\right) \tag{3}$$

$m_s$, $n_s$, and $v_s$ are, respectively, the mass, the number density and the fluid velocity of the species s. $\mathbf{B}_o(0,0,B_o)$ and $c$ are the external static magnetic field and the speed of light in vacuum, respectively.

3 Here the process of dust charging is so swift (typically of the order of 10 seconds than other characteristic processes tens of milliseconds for micro-sized dust grain) than the excitation of acoustic and drift waves. Therefore on a hydrodynamic time scale, the dust charge obtains equilibrium state very quickly. The time, for the formation of modons like nonlinear processes (in the drift wave regime), is much smaller than required for further substantial variations in dust charge, and so dust grain charge will be fixed in each point of the modons.

4 Here in our model, electrons make small gyrations compared to the large size of dust for magnetic field is not very strong and so the change in the dust charge is also minute. In this situation, fast electrons responsible for charging will rush towards the surface of grains along with the direction of $B$ can be treated as Boltzmann. Thus for low frequency dynamics in a magnetized plasma when $\omega/k \ll v_{ts}$ the lighter species can obey Boltzmann distribution. Thus for dust associated frequency regime, in a magnetized plasma when $\omega/k \ll v_{ti}, v_{te}$ (where $v_{ti}$ and $v_{te}$ represent thermal velocity of ions and electrons) the lighter species ($m_{e,i} \to 0$) can obey Boltzmann distribution

$$n_e \simeq n_{e0}(r) \exp\left(\frac{e\phi}{T_e}\right) \simeq n_{e0}(r) \frac{e\phi}{T_e} \tag{4}$$

$$n_i \simeq n_{i0}(r) \exp\left(-\frac{e\phi}{T_i}\right) \simeq -n_{i0}(r) \frac{e\phi}{T_i} \tag{5}$$

in above background density of plasma depends upon r-coordinates.

5 At slow time scale, wavelength $\lambda = 2\pi/k \ll \rho_{g(e,i)} \left[= \left(T_{(e,i)}/m_{(e,i)}\right)^{1/2}/\Omega_{(e,i)}\right]$ where $\rho_{g(e,i)}$ and $\Omega_{(e,i)}$ is the gyroradius and gyrofrequency of electrons and ions therefore only dust grains are taken to be magnetized while magnetization of lighter species $(e,i)$ will be ignored.

6. Many observations from space and in the laboratory plasma indicate deviation of Maxwellian distribution of particles to non- Maxwellian. Various process are responsible to complicate the flows of particles such as turbulence, electrons acceleration, reconnection, external force fields, etc. At low energies, the distribution appears reasonably Maxwellian but have "superthermal tail" at high energies. Therefore we will also discuss a case where consideration of non-Maxwellain distributed lighter species (electrons and ions in this case) will be taken into account for the formation of steady state vortex structures. We aim to adopt kappa and Cairns distribution, normalized number density expressions are given by

$$n_{e(i)} = \left[1 \mp \left(\kappa - \frac{3}{2}\right)^{-1} \frac{\phi}{\sigma}\right]^{-\kappa+1/2} \quad (6)$$

and

$$n_{e(i)} = \left[1 \mp \frac{\beta}{\sigma}\phi \pm \frac{\beta}{\sigma^2}\phi^2\right] e^{\pm\phi} \quad (7)$$

where $\kappa$ is the spectral index measuring the deviation form Maxwellian distribution, $\sigma = T_i/T_e$ and $\beta = 4\alpha/(1+3\alpha)$. For the electrons in (5), $\sigma = 1$. For $\kappa \to \infty (\alpha = 0)$ Maxwellian distribution is achieved. It is to be noted that (4) and (5) will be used only when dust is active in slow time scale however the first case when dust in background (fast time phenomena) only ions dynamics play the role only Eq. (4) will be sued.

7. While treating this problem, we shall first consider the regime where dust is in background and it is also the fast time process which will be followed by slow time (DAW) dynamics. In the former regime, the dust mass is ignored while in the latter, the dust species is activated.

### III. DUST LADEN PLASMA (FAST TIME SCALE FORMULATION)

Here we investigate linear coupling of dust modified drift (DMD) and dust modified acoustic (DMA) waves at fast time and then subsequent formation of nonlinear structures. For that a three-component cylindrically-bounded inhomogeneous magnetized dusty plasma consisting of electrons, ions and *negatively-charged* dust particles is taken into account where ions are cold $[T_i << T_e]$. For $E = -\nabla\phi$, the equation of motion (Eq. (2)) under the drift approximation $|\partial_t| < \Omega_i$ gives the perpendicular and parallel components of the ion fluid

velocity

$$\mathbf{v}_i \simeq \rho_i^2 \Omega_i \hat{\mathbf{z}} \times \nabla \Phi - \rho_i^2 \frac{d}{dt} \nabla_\perp \Phi$$

$$= \mathbf{v}_E + \mathbf{v}_{pi} \qquad (8)$$

where $v_E$ and $v_{pi}$ represent the $\mathbf{E} \times \mathbf{B}$ and ion polarization drift, respectively and $\Phi = e\phi/T_e$ is the normalized potential and $\rho_i = \frac{(T_e/m_i)^{1/2}}{\Omega_i}$ written in the dimensions of length. Parallel equation of ions is

$$\left(\frac{\partial}{\partial t} + v_0 \frac{\partial}{\partial z} + v_E \cdot \nabla\right) v_{iz} + \frac{c}{B_0} \hat{\mathbf{z}} \times \nabla v_0 . \nabla \phi + \frac{e}{m_i} \frac{\partial \phi}{\partial z} = 0 \qquad (9)$$

Above equation has been derived using Eq. (2) where we have used

$$\frac{d}{dt} = \left(\frac{\partial}{\partial t} + v_0 \cdot \nabla + v_E \cdot \nabla\right) \qquad (10)$$

and $\frac{|\mathbf{v}_{pi}|}{|\mathbf{v}_E|} \simeq O(\epsilon)$ where $\epsilon < 1$. The continuity equations of ions in the linear limit reads as:

$$\frac{\partial n_i}{\partial t} + \nabla \cdot \{n_i \mathbf{v}_E + n_i \mathbf{v}_{pi}\} + \frac{\partial}{\partial z}(n_i v_{iz}) = 0 \qquad (11)$$

here for the dust modified perturbations at the slow time scale compared to ions, electron inertia is neglected ($m_e \to 0$) so we obtain Boltzmann density distribution

$$n_e \simeq n_{e0}(r) e^\Phi \simeq n_{e0}(r) \Phi \qquad (12)$$

Poisson equation in this case (fast time) where dust in inactive reads as:

$$n_{i1} \simeq n_{e0} \Phi - \frac{T_e}{4\pi e^2} \nabla^2 \Phi \qquad (13)$$

Substitution of $v_E$ and $v_{pi}$ from Eq. (8) in Eq. (11) and writing down the normalized potential returns us:

$$\left(\frac{\partial}{\partial t} + v_0 \frac{\partial}{\partial z} + \frac{c_{si}^2}{\Omega_i} \hat{\mathbf{z}} \times \nabla \Phi \cdot \nabla\right) v_{iz} + \frac{c_{si}^2}{\Omega_i} \hat{\mathbf{z}} \times \nabla v_0 \cdot \nabla \Phi + c_{si}^2 \frac{\partial \Phi}{\partial z} = 0 \qquad (14)$$

Representation of $\hat{\mathbf{z}} \times \nabla \Phi \cdot \nabla v_{iz}$ and $\hat{\mathbf{z}} \times \nabla v_0 \cdot \nabla \Phi$ as given in the appendix gives us

$$\left(\frac{\partial}{\partial t} + v_0 \frac{\partial}{\partial z}\right) v_{iz} + \left(\frac{n_{e0}}{n_{i0}}\right) \frac{c_{si}^2}{\Omega_i} \{\Phi, \mathbf{v}_{iz}\} + \left(\frac{n_{e0}}{n_{i0}}\right) \frac{c_{si}^2}{\Omega_i} \frac{1}{r} \frac{\partial \Phi}{\partial \theta} \frac{\partial v_0}{\partial r} + \left(\frac{n_{e0}}{n_{i0}}\right) c_{si}^2 \frac{\partial \Phi}{\partial z} = \mathbf{0} \qquad (15)$$

where $\{\Phi, \mathbf{v}_{iz}\}$ is the Poisson's bracket and $c_{si} = (n_{i0}/n_{e0} T_e/m_i)^{1/2}$ is the dust modified ion sound velocity.

## A. Coupling of DMD and DMA waves

Now taking equation of motion in the following form:

$$\left(\frac{\partial}{\partial t} + v_0 \frac{\partial}{\partial z} + v_E \cdot \nabla\right) n_{i1} + \frac{cT_e}{eB_0}\hat{\mathbf{z}} \times \nabla n_{i0}.\nabla\Phi - \frac{n_{i0}c}{B_0\Omega_i}\frac{d}{dt}\nabla\Phi + n_{i0}\frac{\partial v_{iz}}{\partial z} = 0 \quad (16)$$

substitution of $v_E$, from (8) and from (15) gives

$$\left[1 + \frac{Z_d n_{d0}}{n_{e0}}\left(\lambda_{DD}^2 + \frac{n_{e0}^2}{n_{i0}Z_d n_{d0}}\frac{c_{si}^2}{\Omega_i^2}\right)\nabla_\perp^2\right]\frac{\partial\Phi}{\partial t} - v_0\frac{\partial\Phi}{\partial z} -$$

$$\frac{c_{si}^2}{\Omega_i}\frac{1}{r}\left(\ln n_{e0}\frac{\partial\Phi}{\partial\theta}\right) + \frac{n_{e0}}{n_{i0}}\frac{c_{si}^2}{\Omega_i^3}\left(1 + \frac{Z_d n_{d0}}{n_{e0}}\lambda_{DD}^2\Omega_i^2\right)\{\Phi, \nabla^2\Phi\} - \frac{\partial v_{iz}}{\partial z} = 0 \quad (17)$$

where $\lambda_{DD} = (T_e/4\pi Z_d n_{d0}e^2)^{\frac{1}{2}}$ is the Debye length. Equations (16) and (17) are the two coupled nonlinear equations in $\Phi$ and $V$. $\{\Phi, \nabla^2\Phi\}$ is the convective ion polarization drift.

Now we obtain linear dispersion relation for the coupling of dust modified ion acoustic and drift waves,. Here plasma particles are assumed to flowing along the axial direction with linear velocity gradient along radial coordinate as given by

$$v_0(r) = s + qr^2 \quad (18)$$

where $s$ and $q$ are arbitrary constants. Also we suppose density profiles of all plasma particles follow the Gaussian distribution

$$n_{\mp 0} = n_{\mp 00} \exp\left(\mp\frac{r^2}{r_0^2}\right) \quad (19)$$

where $n_{00}$ is the density of electrons at the axis of the cylinder $r = 0$. it is to be noted that we have written a general expression where $\mp$ denote negative (electrons, dust) and ions respectively, in the next sections, we will need to use this equation again. Now suppose the linear perturbations in the dependent variables $\Theta = n, \Phi, v_{iz}$ be

$$\Theta \equiv \exp i(m\theta + k_z z - \omega t) \quad (20)$$

where wave number $k_\theta = m/r$ in $\theta$ direction and $m$ is the mode number. For the local analysis we consider $\partial/\partial r = 0$. Substitution of above (22) into (16) and (17) gives us:

$$\left[\left\{1 - \frac{Z_d n_{d0}}{n_{e0}}\left(\lambda_{DD}^2 + \frac{n_{e0}^2}{n_{i0}Z_d n_{d0}}\frac{c_{si}^2}{\Omega_i^3}\right)k_\theta^2\right\}(\omega + v_0 k_z) - \frac{2mc_{si}^2}{\Omega_i r_0^2}\right]\Phi + k_z v_{iz} = 0 \quad (21)$$

and

$$v_{iz}(\omega - v_0 k_z) = \left(\frac{n_{e0}}{n_{i0}}\right)\left[c_s^2 k_z + 2qm\frac{c_{si}^2}{\Omega_i}\right]\Phi \tag{22}$$

where $v_0 = \nabla \ln n_{00e} \times e_z = -2(r/r_0^2) e_\theta$ in the cylindrical coordinate system $(r, \theta, z)$. Eqs (21) and (22) lead to the following relation:

$$\omega(\omega - v_0 k_z)\left\{1 - \frac{Z_d n_{d0}}{n_{e0}}\left(\lambda_{DD}^2 + \frac{n_{e0}^2}{n_{i0} Z_d n_{d0}}\frac{c_{si}^2}{\Omega_i^3}\right)k_\theta^2\right\} + (\omega - k_z v_0)\left(k_z v_0 - \frac{2mc_{si}^2}{\Omega_i r_0^2}\right)$$

$$+\frac{n_{e0}}{n_{i0}}\left(\frac{2qmc_{si}^2}{r_0^2 \Omega_i}\right)k_z + \frac{n_{e0}}{n_{i0}}c_{si}^2 k_z^2 = 0 \tag{23}$$

(23) is local linear dispersion relation of coupled DMD and DMAwaves in a cylindrical plasma having flow along $z$ dimension and gradient of density is in radial direction.

In case of $v_0 = 0$, it reduces to a simple quadratic equation

$$\omega^2\left\{1 - \frac{Z_d n_{d0}}{n_{e0}}\left(\lambda_{DD}^2 + \frac{n_{e0}^2}{n_{i0} Z_d n_{d0}}\frac{c_{si}^2}{\Omega_i^3}\right)k_\theta^2\right\} - \frac{2mc_{si}^2}{\Omega_i r_0^2}\omega + \frac{n_{e0}}{n_{i0}}c_{si}^2 k_z^2 = 0 \tag{24}$$

Roots of above are given as:

$$\omega = \frac{\omega_e^* \pm \sqrt{\omega_e^{*2} - 4\frac{n_{e0}}{n_{i0}}\left[1 - \frac{Z_d n_{d0}}{n_{e0}}\left(\lambda_{DD}^2 + \frac{n_{e0}^2}{n_{i0} Z_d n_{d0}}\frac{c_{si}^2}{\Omega_i^3}\right)k_\theta^2\right]c_{si}^2 k_z^2}}{2 - \frac{2Z_d n_{d0}}{n_{e0}}\left(\lambda_{DD}^2 + \frac{n_{e0}^2}{n_{i0} Z_d n_{d0}}\frac{c_{si}^2}{\Omega_i^3}\right)k_\theta^2} \tag{25}$$

where $\omega_e^* = 2mc_{si}^2/\Omega_i r_0^2$ is the modified drift frequency. We can transform above relation in Cartesian geometry, for that we express $k_\theta = k_y$ and

$$n_{e0}(x) = n_{00}\exp\left(-\frac{x}{L_n}\right) \tag{26}$$

where $L_n = \left|\frac{1}{n_{e0}}\frac{dn_{e0}}{dx}\right|^{-1}$ represents scale length of the density gradients.

$$\omega^2\left\{1 - \frac{Z_d n_{d0}}{n_{e0}}\left(\lambda_{DD}^2 + \frac{n_{e0}^2}{n_{i0} Z_d n_{d0}}\frac{c_{si}^2}{\Omega_i^3}\right)k_y^2\right\} - \omega_e^*\omega + \frac{n_{e0}}{n_{i0}}c_{si}^2 k_z^2 = 0 \tag{27}$$

where $\omega_e^* = \left(1 + \frac{Z_d n_{d0}}{n_{e0}}\right)\frac{cT_e}{eB_0 n_{e0}}\frac{dn_{e0}}{dx}k_y$ is the modified drift frequency due to the presence of dust and is much larger than the usual drift velocity in electron ion plasma.

### B. Nonlinear analysis and harmonics of modons

When the velocity of two dimensional fluid affiliated with dispersive waves gains locally larger value than the phase velocity because of nonlinear effects, then the wave shows curving of wave front which finally leads to the creation of vortex structure. Meiss and Horton [38] studied one particular group of solutions for drift modons (dipolar vortices) described by coupled vortices of both positive and negative potential. Whereas global vortex solutions which are another class of solution have been explored by M. Yu [39] for Hasegawa Mima (HM) equation and by Nycander [40] for flute modes in inhomogenous plasma. From mathematical point of view these two type of vortex structures differ from one another and difference arises due to the boundary conditions we use. Recenlty 3D soliton drift wave (3D modon) and their collisions with the aid of numerical simulations were invetigated[41]. Modons are assumed to propagate in the $(\theta, z)$ directions so the moving coordinate is defines as: $\eta = \theta + \alpha_0 z - \Omega^* t$, where $\Omega^*$ is the constant rotation frequency of the modons. $\alpha_0$ is angle along the $z-axis$.

$$\frac{\partial}{\partial t} = -\Omega^* \frac{\partial}{\partial \eta}; \frac{\partial}{\partial z} = \alpha_0 \frac{\partial}{\partial \eta}; \frac{\partial}{\partial \theta} = \frac{\partial}{\partial \eta} \tag{28}$$

Substitution of transformation given in Eq. (28) into (9) and assuming $|\alpha_0 \partial_z| << \Omega^* |\partial_t|$, we obtain

$$\left[\frac{\partial}{\partial \eta} - \frac{n_{e0}}{n_{i0}} \frac{c_{si}^2}{\Omega^* \Omega_i} \frac{1}{r} \left(\frac{\partial \Phi}{\partial r} \frac{\partial}{\partial \eta} - \frac{\partial \Phi}{\partial \eta} \frac{\partial}{\partial r}\right)\right] v_{iz} = \frac{n_{e0}}{n_{i0}} \frac{c_{si}^2}{\Omega^*} \left(\alpha_0 + \frac{2q}{\Omega_i}\right) \frac{\partial \Phi}{\partial \eta} \tag{29}$$

here we define an operator as:

$$\widehat{D}_\phi = \frac{1}{r}\left(\frac{\partial \Phi}{\partial r}\frac{\partial}{\partial \eta} - \frac{\partial \Phi}{\partial \eta}\frac{\partial}{\partial r}\right) \tag{30}$$

and rearranging above (28), one gets

$$\left(\frac{\partial}{\partial \eta} - \frac{n_{e0}}{n_{i0}} \frac{c_{si}^2}{\Omega^* \Omega_i} \widehat{D}_\phi\right) v_{iz} = \frac{n_{e0}}{n_{i0}} \frac{c_{si}^2}{\Omega^*} \left(\alpha_0 + \frac{2q}{\Omega_i}\right) \frac{\partial \Phi}{\partial \eta} \tag{31}$$

let

$$\widehat{D}_\phi^* = \left(\frac{\partial}{\partial \eta} - \frac{n_{e0}}{n_{i0}} \frac{c_{si}^2}{\Omega^* \Omega_i} \widehat{D}_\phi\right) \tag{32}$$

this enables us to write $\partial \Phi / \partial \eta = \widehat{D}_\phi^* \Phi$, so eq. (31) can be written as

$$\widehat{D}_\phi^* v_{iz} = \frac{n_{e0}}{n_{i0}} \frac{c_{si}^2}{\Omega^*} \left(\alpha_0 + \frac{2q}{\Omega_i}\right) \widehat{D}_\phi^* \Phi \tag{33}$$

which gives us a linear relation between velocity and potential in such form

$$v_{iz} = \frac{n_{e0}}{n_{i0}} \frac{c_{si}^2}{\Omega^*} \left(\alpha_0 + \frac{2q}{\Omega_i}\right) \Phi \tag{34}$$

Transforming to the $\eta$ frame, (9) and substitution of $v_{iz}$ gives us

$$\left[\frac{2}{r_0^2}\frac{c_{si}^2}{\Omega_i} - \alpha_0\left\{\frac{n_{e0}}{n_{i0}}\frac{c_{si}^2}{\Omega^*}\left(\alpha_0 + \frac{2q}{\Omega_i}\right)\right\} - \Omega^*\right]\frac{\partial\Phi}{\partial\eta} - \left\{\Omega^*\frac{Z_d n_{d0}}{n_{e0}}\left(\lambda_{DD}^2 + \frac{n_{e0}^2}{n_{i0}Z_d n_{d0}}\frac{c_{si}^2}{\Omega_i^2}\right)\nabla_\perp^2\right\}\frac{\partial\Phi}{\partial\eta}$$

$$+\frac{n_{e0}}{n_{i0}}\frac{c_{si}^2}{\Omega_i^3}\left(1 + \frac{Z_d n_{d0}}{n_{e0}}\lambda_{DD}^2\Omega_i^2\right)\widehat{D}_\phi\nabla^2\Phi = 0 \tag{35}$$

Above is Hasegawa-Mima like equation modified by the presence of the dust in background. This equation provides the provides the profile of the electrostatic potential of the drift waves in an inhomogeneous dusty plasma. Now to find the solution of above, we simplify the above equation in such form

$$\left[\frac{2}{r_0^2}\frac{c_{si}^2}{\Omega_i} - \frac{n_{e0}}{n_{i0}}\frac{\alpha_0 c_{si}^2}{\Omega^*}\left(\alpha_0 + \frac{2q}{\Omega_i}\right) - \Omega^*\right]\frac{\partial\Phi}{\partial\eta} = X\left[\frac{\partial}{\partial\eta} - Y\widehat{D}_\phi\right]\nabla^2\Phi \tag{36}$$

where

$$X = \frac{Z_d n_{d0}\Omega^*}{n_{e0}}\left(\lambda_{DD}^2 + \frac{n_{e0}^2}{n_{i0}Z_d n_{d0}}\frac{c_{si}^2}{\Omega_i^2}\right) \tag{37}$$

and

$$Y = \frac{n_{e0}^2}{Z_d n_{d0} n_{i0}}\frac{c_{si}^2}{\Omega^*\Omega_i^3}\frac{\left(1 + \frac{Z_d n_{d0}}{n_{e0}}\lambda_{DD}^2\Omega_i^2\right)}{\left(\lambda_{DD}^2 + \frac{n_{e0}^2}{n_{i0}Z_d n_{d0}}\frac{c_{si}^2}{\Omega_i^2}\right)} \tag{38}$$

Now we introduce another operator

$$L_\phi = \frac{\partial}{\partial\eta} - Y\widehat{D}_\phi \tag{39}$$

then Eq. (36) can be expressed as

$$L_\phi\left[\nabla^2\Phi + \frac{1}{X}\left\{\Omega^* + \frac{n_{e0}}{n_{i0}}\frac{\alpha_0 c_{si}^2}{\Omega^*}\left(\alpha_0 + \frac{2q}{\Omega_i}\right) - \frac{2}{r_0^2}\frac{c_{si}^2}{\Omega_i}\right\}\Phi\right] = 0 \tag{40}$$

or

$$\nabla^2\Phi + \Lambda^2\Phi = Ar^2 \tag{41}$$

where

$$\Lambda = \frac{n_{e0}n_{i0}\Omega_i^2}{c_{si}^2(n_{e0}^2 + \lambda_{DD}^2 n_{i0}Z_d n_{d0})}\left\{1 + \frac{n_{e0}}{n_{i0}}\frac{\alpha_0 c_{si}^2}{\Omega^{*2}}\left(\alpha_0 + \frac{2q}{\Omega_i}\right) - \frac{2}{r_0^2\Omega^*}\frac{c_{si}^2}{\Omega_i}\right\} \tag{42}$$

and $\Lambda$ some constant. The general solution of (41) is

$$\Phi(r,\eta) = \Phi_m J_n(r\Lambda)cos(n\eta) + \frac{Ar^2}{\Lambda^2} - \frac{4A}{\Lambda^4} \tag{43}$$

where $\Phi_m$ is maximum amplitude of vortex, $J_n$ the Bessel function of order $n$. Solution (43) must satisfy the boundary conditions that radial fluid velocity vanishes at $r = 0$ and $r = r_0$.

$\Phi = 0$ at $r = r_0$, we require $\Lambda = k_a/r_0$ where $k_a$ are zeros of Bessel function and $a = 1, 2, 3....$ For the condition $r = 0$ to satisfy, order bessel function "$n$" should be $\geqslant 1$. For $\Phi = 0$ at $r = r_0$ solution of (43) is given as:

$$\Phi(r, \eta) = \Phi_m J_n\left(\frac{k_a r}{r_0}\right) cos(n\eta) + \frac{A_0^2 r_0^2}{k_a^2}\left(r^2 - \frac{4r_0^2}{k_\alpha^2}\right) \tag{44}$$

where

$$A_0 = \frac{n_{e0} n_{i0} \Omega_i^2}{c_{si}^2 \left(n_{e0}^2 + \lambda_{DD}^2 n_{i0} Z_d n_{d0}\right)} \left\{1 + \frac{n_{e0}}{n_{i0}} \frac{\alpha_0 c_{si}^2}{\Omega^{*2}} \left(\alpha_0 + \frac{2q}{\Omega_i}\right) - \frac{2}{r_0^2 \Omega^*} \frac{c_{si}^2}{\Omega_i}\right\} - \frac{k_a^2}{r_0^2} \tag{45}$$

here this solution shows formation of modons which may appear in many harmonics of the azimuthal angle ($n = 1, 2, 3...$). Eq. (44) clearly indicated, modon depends upon the number density of the plasma particles, magnetic field and rotation frequency. The azimuthal wave numbers ($n$) determines the number of modon pairs. The presents work does not deal with the stability of the vortex however this solution is more general and physically realistic. Above equation shows, profile of the modons in cylindrical dusty plasma depends upon rotational frequency, number dencity and radius of the cylinder.

### IV. DUST LADEN PLASMA (SLOW TIME SCALE FORMULATION)

Here we assume a situation where the time required for the variation in the velocity and density of electrons/ions is much shorter than required for the change in these parameters of dust, i.e.,

$$v_e \left(\frac{\partial v_e}{\partial t}\right)^{-1}, n_e \left(\frac{\partial n_e}{\partial t}\right)^{-1} < t_i \sim \frac{1}{\omega_{pi}} << t_d \sim \frac{1}{\omega_{pd}} \tag{46}$$

where $\omega_{pi(p)}$ is the Langmuir frequencies for ion (dust). Therefore here dust particles are considered in motion whereas dynamics of lighter electrons and ions have been ignored, whereas the neutrality condition ($n_{io} = Z_d n_{do} + n_{eo}$) holds. All three components of the plasma ($s = i, e$ and negative $d$) are taken to have same sheared flow $v_{d0} = v_{i0} = v_{e0} = v_0(r)\hat{\mathbf{z}}$, therefore $B_0$ is taken to be constant.

Generally we know dusty plasmas are low temperature systems and therefore $T_d << T_i, T_e$. Momentum equation for the dust under the condition of drift approximation $|\frac{\partial}{\partial t}| < \Omega_d$ gives

$$\mathbf{v}_d \simeq \frac{c}{B_0}\hat{\mathbf{z}} \times \nabla\phi + \frac{c}{B_0 \Omega_d} \frac{d}{dt} \nabla_\perp \phi = \mathbf{v}_E + \mathbf{v}_{pd} \tag{47}$$

where $v_E$ and $v_{pd}$ represent the electric and polarization drift of the negatively charged dust, respectively. $\Omega_d = eB_0/m_d c$ is the dust gyro frequency whereas $|\mathbf{v}_{pd}|/|\mathbf{v}_E| \simeq O(\epsilon)$ where $\epsilon < 1$. Continuity equation for dust particles is given as

$$\frac{\partial n_d}{\partial t} + \nabla \cdot \{n_d \mathbf{v}_E + n_d \mathbf{v}_{pd}\} + \frac{\partial}{\partial z}(n_d v_{dz}) = 0 \tag{48}$$

Here we will ignore the charge fluctuation effect which is an important feature of the dusty plasma because variations in the charge cause dissipation affects leading to the damping of the relevant acoustic mode, Moreover to incorporate variable dust charge affect one needs to deal with the complex function of charging cross section which depends upon parameters like external magnetic field and the impact of the particles approaching grains to distances smaller than the particle size.

Here we assume gyroradius of electrons is much smaller than the size of dust particles and so changes in the dust charge can be negligible, in such case, electrons reach quite fast the surface of dust practices along the direction of magnetic field ( $B$) and therefore electrons move faster due to higher mobility taking place in the dust charging process.. It should be mentioned that in our analysis, the occurrence time for the formation of vortex like structure is much shorter than the time needed for further significant variation in dust charge, and therefore we do not taken into account dust charge fluctuations which is indeed an important feature of dusty plasmas.

Poisson equation is given as

$$\nabla^2 \phi = 4\pi e \left( Z_d n_d + n_{e0}(r)\frac{e\phi}{T_e} - n_{i0}(r)\frac{e\phi}{T_i} \right) \tag{49}$$

From Eqs. (47-49) and introducing normalized potential $\Phi = e\phi/T_e$, one gets,

$$\left(\frac{\partial}{\partial t} + v_0 \frac{\partial}{\partial z} + \frac{c}{B_0}\hat{\mathbf{z}} \times \nabla\phi \cdot \nabla\right)\left(\frac{T_i}{4\pi e^2}\nabla^2 - N_D \Phi\right) + \frac{Z_d c T_i}{eB_0}\hat{\mathbf{z}} \times \nabla\Phi \cdot \nabla n_{d0}$$

$$+\frac{Z_d n_{d0} c T_i}{eB_0 \Omega_d}\left(\frac{\partial}{\partial t} + v_0\frac{\partial}{\partial z} + \frac{cT_i}{eB_0}\hat{\mathbf{z}} \times \nabla\Phi \cdot \nabla\right)\nabla^2\Phi + Z_d n_{d0}\frac{\partial v_{dz}}{\partial z} = 0 \tag{50}$$

to obtain above equation, higher order terms have been ignored and $N_D = (T_i n_{e0} + n_{i0} T_e)/T_e$, $\rho_d = (c_{sd}/\Omega_d)(Z_d N_D)^{-1/2}$. The above equation can be expressed as

$$\left(\frac{\partial}{\partial t} + v_0 \frac{\partial}{\partial z}\right)\mathbf{V}_{dz} + \Omega_d \rho_d^2 \frac{cT_e}{eB_0}\{\Phi, V_{dz}\} + \Omega_d \rho_d^2 \frac{1}{r}\frac{\partial \Phi}{\partial \theta}\frac{\partial v_0}{\partial r} = \frac{c_{sd}^2}{N_d}\frac{\partial \Phi}{\partial z} \tag{51}$$

Continuity equation can be written as:

$$\left(-N_D + \lambda_{DD}^2 \nabla_\perp^2 + \rho_d^2 \nabla^2\right) \frac{\partial \Phi}{\partial t} - v_0 N_D \frac{\partial \Phi}{\partial z} + \Omega_d \rho_d^2 (\lambda_{DD}^2 + \rho_d^2)\{\Phi, \nabla^2 \Phi\} + V_{*n} \frac{\partial \Phi}{\partial \theta} + \frac{\partial V_{dz}}{\partial z} = 0 \quad (52)$$

where $V_{*n} = \Omega_d \rho_d^2 \frac{1}{r} \frac{\partial}{\partial r} \ln n_{d0}$ and $\{\Phi, \mathbf{V}_{dz}\} = \frac{1}{r}(\partial_r \Phi \partial_\theta \mathbf{V}_{dz} - \partial_\theta \Phi \partial_r \mathbf{V}_{dz})$ the Poisson bracket.

### A. Linear dispersion relation of the coupled UDD and UDA waves

Proceeding along the lines of previous calculations for dispersion relation, we obtain a dispersion relation for the DLH wave given by

$$\omega(\omega - v_0 k_z)\left\{N_d + \left(\lambda_{DD}^2 + \rho_d^2\right) k_\theta^2\right\} - (\omega - k_z v_0)\left(N_d k_z v_0 + \frac{2m\Omega_d \rho_d^2}{r_0^2}\right) - \left[\left(\frac{2qm\Omega_d \rho_d^2}{r_0^2}\right) + \frac{c_{sd}^2}{N_D} k_z\right] k_z = 0 \quad (53)$$

and when $v_0 = 0$, above dispersion relation is reduced to

$$\omega^2 \left\{N_d + \left(\lambda_{DD}^2 + \rho_d^2\right) k_\theta^2\right\} - \omega_0 \omega_d^* - \frac{c_{sd}^2}{N_D} k_z^2 = 0 \quad (54)$$

and now $\omega_d^* = 2m\Omega_d \rho_d^2 / r_0^2$. Roots of above equation are:

$$\omega = \frac{\omega_d^* \pm \sqrt{\omega_d^{*2} - 4 \frac{c_{sd}^2}{N_D}\left\{N_d + (\lambda_{DD}^2 + \rho_d^2) k_\theta^2\right\} k_z^2}}{2\left\{N_d + (\lambda_{DD}^2 + \rho_d^2) k_\theta^2\right\}} \quad (55)$$

And in Cartesian system

$$\omega^2 \left\{N_d + \left(\lambda_{DD}^2 + \rho_d^2\right) k_y^2\right\} - \omega_0 \omega_d^* - \frac{c_{sd}^2}{N_D} k_z^2 = 0 \quad (56)$$

$\omega_d^* = \Omega_d \rho_d^2 \hat{\mathbf{z}} \times \nabla \ln n_{d0} \cdot \mathbf{k}_\perp$.

### B. Formation of vortex like structures

In this case we assume ultra low frequency modon is rotating with $\Omega_0$. Proceeding along the lines of previous calculations given in Sec. (III), for a moving coordinate $\eta = \theta + \alpha_0 z - \Omega_0 t$ and $|\alpha_0 \partial_z| << \Omega_0 |\partial_t|$, we obtain

$$\left(\frac{\partial}{\partial \eta} - \frac{\rho_d^2 \Omega_d}{\Omega_0} D_\phi\right) V_{dz} = \left(\frac{2b\Omega_d \rho_d^2 - \alpha_0 c_{sd}^2}{\Omega_0}\right) \frac{\partial \Phi}{\partial \eta} \quad (57)$$

where $D_\phi = \frac{1}{r}\left(\frac{\partial \Phi}{\partial r} \frac{\partial}{\partial \eta} - \frac{\partial \Phi}{\partial \eta} \frac{\partial}{\partial r}\right)$. Following the algebraic step given in the previous case, we get

$$V_{dz} = \left\{\frac{(2b\Omega_d \rho_d^2 - \alpha_0 c_{sd}^2)}{\Omega_0}\right\} \Phi \quad (58)$$

Transforming to the $\eta$ frame gives us:

$$\Omega_0 \left( \left(-N_D + \rho_d^2 \nabla^2 + \lambda_{DD}^2 \nabla^2 + \frac{V_{*n}}{\Omega_0}\right) \right) \frac{\partial \Phi}{\partial \eta} = \Omega_d \rho_d^2 \left(\lambda_{DD}^2 + \rho_d^2\right)$$

$$\times \frac{1}{r} \left( \frac{\partial \Phi}{\partial r} \frac{\partial}{\partial \eta} \nabla^2 \Phi - \frac{\partial \Phi}{\partial \eta} \frac{\partial}{\partial r} \nabla^2 \Phi \right) - \alpha_0 \left( \frac{2b\Omega_d \rho_d^2 - \alpha_0 c_{sd}^2}{\Omega_0^2} \right) \frac{\partial \Phi}{\partial \eta} \qquad (59)$$

Introducing $L_\phi = \left( \frac{\partial}{\partial \eta} - \frac{\Omega_d \rho_d^2}{\Omega_0} D_\phi \right)$ as another operator in above equation gives us:

$$L_\phi \left[ \nabla^2 \Phi + \left\{ \left( \frac{V_{*n}}{\Omega_0} - \frac{\alpha_0 (2b\Omega_d \rho_d^2 - c_{sd}^2 \alpha_0)}{\Omega_0^2} - N_D \right) \left(\lambda_{DD}^2 + \rho_d^2\right)^{-1} \right\} \Phi \right] = 0 \qquad (60)$$

or

$$\nabla^2 \Phi + \chi^2 \Phi = A r^2 \qquad (61)$$

where $A = \frac{B}{2H}$ and

$$\chi^2 = \left(\lambda_{DD}^2 + \rho_d^2\right)^{-1} \left( \frac{V_{*n}}{\Omega_0} - \frac{\alpha_0 (2b\Omega_d \rho_d^2 - c_d^2 \alpha_0)}{\Omega_0^2} - N_D \right) - B \qquad (62)$$

Solution of (61) is

$$\Phi(r, \eta) = \Phi_{\max} J_n(\chi r) cos(n\eta) + \frac{A r^2}{\chi^2} - \frac{4A}{\chi^4} \qquad (63)$$

where maximum amplitude of modon is given by $\Phi_{\max}$. Here $\chi = k_a/r_0$ where $k_a$ are zeros of Bessel function and solution:

$$\Phi(r, \eta) = \Phi_{\max} J_n\left(\frac{k_a r}{r_0}\right) cos(n\eta) + \frac{A^2 r_0^2}{k_a^2}\left(r^2 - \frac{4 r_0^2}{k_\alpha^2}\right) \qquad (64)$$

where

$$A = \frac{1}{2H} \left[ \frac{\frac{V_{*n}}{\Omega_0} - \frac{\alpha_0(2b\Omega_d \rho_d^2 - c_d^2 \alpha_0)}{\Omega_0^2} - N_D}{(\lambda_{DD}^2 + \rho_d^2)} - \frac{k_a^2}{r_0^2} \right] \qquad (65)$$

Here in $A$ density factor is added as a negative term unlike previous case when dust was only in background whereas again the structure of the vortex depends upon magnetic field and rotaion frequency.

## V. COMET HALLEY PLASMA

As stated in the introduction we consider a comet Halley plasma whose components are negative dust, electrons, and singly ionized negatively and positively charged ions. The condition of charge neutrality at equilibrium gives $n_{p0} = n_{e0} - Z_d n_{d0} + n_{n0}$, where now ($n_p$, $n_n$) represent hydroxides (OH$^+$,OH$^-$),hydrogen (H$^+$, H$^-$), oxygen (O$^+$, O$^-$), silicon (Si$^+$, Si$^-$) etc. observed by the Vega I and II spacecraft. Here we assume $m_n(T_n) \equiv m_p(T_p)$.

In most of the astrophysical and lab dusty plasmas, electrons can stick to the surface of dust after a collision and get absorbed due to their high mobility and so their density reduces. Such a system can be treated as electron depleted plasma because now there is strong deficiency of electrons, this is equivalent to the assumption $n_{oe}/n_{op} \ll T_e/T_p$ and $\omega \sim \Omega_{(p,n)}$. In this scenario plasma oscillations are below the electron Debye length therefore role of ions become more significant for the excitation of a very low frequency wave (such as dust drift waves). Thus, the quasi-neutrality condition now reads as $\delta n_p \simeq \delta n_n$. In the comet Halley plasma role of negative ions becomes more crucial. Here we assume ions to be cold and components of wave vector $k$ along and perpendicular to magnetic field are denoted by $k_z$ and $k_y$ i.e. $\nabla = (0, k_y, k_z)$.

Vector product of equation of motion (for negative and positive ions) with $\hat{z}$ gives us perpendicular velocity

$$v_{(p,n)\perp} = \frac{1}{B_0}\hat{z} \times \nabla \phi \mp \frac{1}{\Omega_i}\left(\frac{1}{B_0}(\hat{z} \times \nabla \phi) \times \hat{z}\right) \tag{66}$$

which upon further simplification gives us

$$v_{(p,n)\perp} = \frac{ik_y \phi}{B_0}\hat{x} \mp \frac{k_y \omega}{\Omega_{(p,n)}B_0}\phi \hat{y} \tag{67}$$

Parallel equation of motion and continuity equation for both species of ions give us

$$\frac{n_{p1}}{n_{p0}} = \left(\frac{ek_z^2}{m_p \omega^2} - \frac{k_y^2}{\Omega_p B_0}\right)\phi \tag{68}$$

$$\frac{n_{n1}}{n_{n0}} = \left(-\frac{ek_z^2}{m_n \omega^2} + \frac{k_y^2}{\Omega_n B_0}\right)\phi \tag{69}$$

where $|\Omega_p| = |\Omega_n| = |\Omega_{pn}|$. Substitution of above in the quasi neutrality condition and using the neutrality condition $n_{p0} = n_{e0} - Z_d n_{d0} + n_{n0}$, we obtain the dispersion relation

$$\omega^2 = \frac{k_z^2}{k_y^2}\Omega_{pn}^2 \tag{70}$$

for the dust drift mode excited in the comet Halley plasma. In above equation (70), there is no contribution of dust and electrons which are in background which was unexpected as the presence of dust even in background has modified the acoustic and drift waves in the previous cases. It is obvious from (70), that contribution of electric drifts and gradients has been vanished. The high frequency $E \times B$ currents for the both ions have same magnitude but different sign so are canceled out thus pay no contribution to the net current

$$\mathbf{j}_p = \frac{n_{op}q^3 E \times B}{m_p^2 c(\Omega_{pn}^2 - \omega^2)} \tag{71}$$

$$\mathbf{j}_n = \frac{n_{on}q^3 E \times B}{m_n^2 c(\Omega_{pn}^2 - \omega^2)} \tag{72}$$

Polarization currents due to hydroxides (OH$^+$,OH$^-$),hydrogen (H$^+$, H$^-$), oxygen (O$^+$, O$^-$) and silicon (Si$^+$, Si$^-$) remain which have now been added. Ions polarization drift is responsible for this mode which is charge-dependent. This looks like *convective cell* mode of fluid mechanics. In usual plasma counterpart of this mode is the Alfvén waves where dynamics is governed by the polarization drift of ions and electric drift of both the electrons and ions is cancelled out. This mode seems to appear very briefly, how stable is this mode in comet Halley or electronegative plasma can be studied however that is not scope of this manuscript but we plan to investigate that with kinetic treatment as well in future papers.

Here we investigate linear and nonlinear dynamics of this mode. For this we assume $v_{0p} = v_{0n} = v_0(r)\hat{\mathbf{z}}$. In this case we will write equation of motion and continuity for the ions and obviously $|\partial_t| < \Omega_{p(n)}$ which are:

$$\mathbf{v}_{\perp_{p(n)}} = \frac{c}{B_0}\hat{\mathbf{z}} \times \nabla \phi \mp \frac{c}{B_0 \Omega_i} d_t \nabla_\perp \phi = \mathbf{v}_E \mp \mathbf{v}_P) \tag{73}$$

The continuity equations are

$$\frac{\partial n_{p(n)}}{\partial t} + \nabla \cdot \left\{ n_{p(n)} \mathbf{v}_E + n_{p(n)} \mathbf{v}_{pd} \right\} + \frac{\partial}{\partial z}(n_{p(n)} v_{p(n)z}) = 0 \tag{74}$$

We express (74) in the following form:

$$\frac{\partial n_{p(n)}}{\partial t} + n_{p(n)0}\left[ v_0 \frac{\partial}{\partial z}\left(\frac{n_{p(n)}}{n_{p(n)0}}\right) - \frac{c}{B_0}\frac{1}{r}\left(\frac{\partial}{\partial r} ln n_{p(n)0}\right)\frac{\partial \Phi}{\partial \theta} \mp \frac{c}{B_0 \Omega_{p(n)}}\frac{\partial}{\partial t}\nabla_\perp^2 \phi \mp \frac{c^2}{B_0^2 \Omega_{p(n)}}\{\phi, \nabla_\perp^2 \phi\} + \frac{\partial v_{zp(n)}}{\partial z}\right] = \tag{75}$$

## VI.  LINEAR SOLUTION

Continuity equations for $+ve$ and $-ve$ ions become, respectively

$$\frac{n_{p(n)}}{n_{p(n)0}} = \frac{1}{(\omega - v_0 k_z)}\left[\frac{2m_p c_s^2}{r_0^2 \Omega_{pn}}\right]\Psi \pm \frac{c_s^2}{\Omega_{pn}^2}F \pm \left[\frac{\nu_{tpn}^2 \Omega_{pn} k_z^2 \mp 2mbk_z c_s^2}{(\omega - v_0 k_z)^2 \Omega_{pn}}\right]\Psi \tag{76}$$

where we have defined $\Psi = e\phi/T_{pn}$ where $T_n(m_n) = T_p(m_p) = T_{pn}(m_{pn})$. Using quasi-neutrality condition in (75) and (76) yields

$$\left[\frac{2mc_s^2}{r_0^2\Omega_{pn}} - \frac{2mbk_zc_s^2}{(\omega - v_0k_z)\Omega_{pn}}\right]\Psi + \frac{n_{p0} + n_{n0}}{(n_{p0} + n_{n0})(\omega - v_0k_z)}\left[\rho_{pn}^2 F + \frac{\nu_{tpn}^2 k_z^2}{(\omega - v_0k_z)^2}\Psi\right] = 0 \quad (77)$$

and where $F = 1/r \partial\Psi/\partial r + \partial^2\Psi/\partial r^2 - (m/r)^2 \Psi$.

or

$$\left[\frac{2mc_s^2}{r_0^2\Omega_{pn}} - \frac{2mbk_zc_s^2}{(\omega - v_0k_z)\Omega_{pn}}\right]\frac{Z_d n_{d0}}{\omega - v_0k_z} + a^1\left[\left(\frac{c_s}{\Omega_{pn}}\right)^2 F + \frac{\nu_{tpn}^2 k_z^2}{(\omega - v_0k_z)^2}\right] = 0 \quad (78)$$

where $a^1 = Z_d n_{d0}\left(1 + \frac{n_{e0}}{Z_d n_{d0}} + 2\frac{n_{n0}}{Z_d n_{d0}}\right)/\left(1 + \frac{Z_d n_{d0}}{n_{e0}}\right)$. Let us define sound like velocity in the comet Halley plasma $c_{s\star}^2 = \left(1 + \frac{Z_d n_{d0}}{n_{e0}}\right)T_{pn}/m_{pm}$, further simplification reduces to

$$a^1\left(\frac{\partial^2\Psi}{\partial r^2} + \frac{1}{r}\frac{\partial\Psi}{\partial r} - \left(\frac{m}{r}\right)^2 \Psi\right)\omega^2 + \left(\frac{2mZ_d n_{d0}}{r_0^2 \Omega_{pn}}\right)\omega\Psi + a^1 k_z^2 - \frac{mbk_z Z_d n_{d0}}{a^1} = 0 \quad (79)$$

. It is obvious from the above relation, in the comet Halley case there is no coupling of acoustic wave which has velocity $c_{s\star}$ and drift waves but it's only convective cell mode which evolves. Above equation can however be written as an ordinary Bessle function which has well-known eigen value solution.

## VII. NONLINEAR SOLUTION

Now we proceed to the investigation of nonlinear dynamics of this so-called convective cell mode excited in the comet Halley plasma and see if this new mode can lead to the formation of modons.

Under the condition $\alpha\partial/\partial_z| << |\Omega_{rot}\partial/\partial_t|$ where $\Omega_{rot}$ is the rotational frequency of the modon. Proceeding along the lines of previous sections, we finally obtain equation which has solution similar to the previous cases giving possibility of modons' formation

$$A^0\frac{\partial\Psi}{\partial\eta} + B^0\frac{\partial\nabla^2}{\partial\eta}\Psi + C^0\{\Psi, \nabla^2\Psi\} = 0 \quad (80)$$

$$A^0 = Z_d n_{d0}\left\{\left(1 + \frac{n_{e0}}{Z_d n_{d0}}\right)\left(\frac{2}{r_0^2}\frac{c_s^2}{\Omega_{pn}^2}\right) + \frac{\alpha^2 v_{ti}^2}{\Omega_{rot}}\left(1 + \frac{n_{e0}}{Z_d n_{d0}} + 2\frac{n_{n0}}{Z_d n_{d0}}\right)\left(1 - \frac{2bc_s^2}{\alpha v_{ti}^2 \Omega_{pn}^2}\right)\right\} \quad (81)$$

$$B^0 = \frac{\Omega_{rot}}{4\pi e}\left\{1 + \left(1 + \frac{n_{e0}}{Z_d n_{d0}} + 2\frac{n_{n0}}{Z_d n_{d0}}\right)\frac{4\pi Z_d e n_{d0} c_s^2 \Omega_{rot}}{\Omega_{pn}^2}\right\} \quad (82)$$

$$C^0 = -\left(1 + \frac{n_{e0}}{Z_d n_{d0}} + 2\frac{n_{n0}}{Z_d n_{d0}}\right)\frac{Z_d n_{d0} c_s^4 \Omega_i}{\Omega_{pn}^4} \quad (83)$$

## VIII. TROPICAL MESOSPHERE PLASMA

Positively charged dust grains due to the thermionic emission of electrons from a fractionally heated meteoroid, as it enters the Earth's atmosphere were studied by Mendis et al. and Sorasio et al.[42, 43] however their treatment was limited to night side of the Earth where role of thermionic emission was considered where photoemission was ignored which is feature of Earth's dayside. Positive dust has also been present in the tokamak like fusion devices where solid particles are produced by plasma-surface interactions. It's important to note that unlike usual dusty plasma systems where number density of electrons is quite less due to the absorption (often treated as electron-depleated plasma), instead for a positive dust, density of electrons is larger than ions following: $(n_{io} + Z_d n_{d0} - n_{eo} - n_{s0} = 0)$. This affects the associated physical processes.

It's useful to mention that during dust charging process, due to primary electron, current flows towards the grain whereas for secondary electron case, current flows out of the positively charged which based is usually given as[44, 45], the orbital motion limited (OML) theory and dust grains are

$$I_e^{\text{sec}} = 3.7\delta_M \pi r_0^2 \sqrt{\frac{8T_e}{\pi m_e}} \left(1 + \frac{eZ_d}{r_0 T_e}\right) \exp eq_d \left(\frac{T_s - T_e}{r_0 T_e T_s}\right) F_{5,B}(x) \quad (84)$$

where temperature $T_s$ is temperature of secondary electrons and $\delta_M$ is the maximum yield which happens when these electrons strike with some kinetic energy $(E_M)$. The function $F_{5,B}(x)$ is given as

$$F_{5,B}(x) = x^2 \int_B^\infty u^2 \exp(-(xu^2 + u))du \quad (85)$$

where $x = E_M/4T_e$ and $B = eZ_d/r_0 T_e x$. In deriving Eq. (84), Boltzmann distributed secondary electrons $n_s \simeq n_{s0}(r)[e\phi/T_s]$ have been taken into account however in more real situations this might not be the case. In our model, we assume that the nonlinear processes under consideration happen after the charging process is stopped. Inclusion of charge variations affect will lead do additional dissipation processes which can impact the modon formation but that is beyond the scope of present study.

In this case linear dispersion relation of coupled drift and acoustic waves turns out to be

$$\omega(\omega-v_0 k_z)\left\{N_{sd}+\left(\lambda_{DD}^2+\rho_d^2\right)k_\theta^2\right\}-(\omega-k_z v_0)\left(N_{sd}k_z v_0-\frac{2m\Omega_d\rho_d^2}{r_0^2}\right)+\left[2qm\Omega_d\rho_d^2+c_{sd}^2 k_z\right]k_z=0 \tag{86}$$

where $N_{sd}=\left(\frac{T_i}{T_e}n_{e0}+n_{i0}+\frac{T_i}{T_s}n_{s0}\right)/Z_d n_{d0}$.

Assuming that density profiles of all plasma particles follow the Gaussian distribution, the resulting nonlinear equation can be written as:

$$L_\phi\left[\nabla^2\Phi-\left\{\left(N_{sd}\frac{V_{*n}}{\Omega_0}-\frac{\alpha_0(2b\Omega_d\rho_d^2-c_{sd}^2\alpha_0)}{\Omega_0^2}-\right)\left(\lambda_{DD}^2+\rho_d^2\right)^{-1}\right\}\Phi\right]=0 \tag{87}$$

above can be expressed like $\nabla^2\Phi+\chi_s^2\Phi=A_{os}r^2$ whose solution can be given as:

$$\Phi(r,\eta)=\Phi_{\max}J_n\left(\frac{k_a r}{r_0}\right)cos(n\eta)+\frac{A_{os}^2 r_0^2}{k_a^2}\left(r^2-\frac{4r_0^2}{k_\alpha^2}\right) \tag{88}$$

where

$$A_{os}=\frac{1}{2H}\left[\frac{\frac{\alpha_0(2b\Omega_d\rho_d^2-c_d^2\alpha_0)}{\Omega_0^2}-\frac{2\Omega_d\rho_d^2}{\Omega_0 r_0^2}-N_{sd}}{(\lambda_{DD}^2+\rho_d^2)}-\frac{k_a^2}{r_0^2}\right] \tag{89}$$

## IX. KAPPA AND CAIRNS DISTRIBUTION

In this section, we will develop nonlinear equation admitting vortex like solutions for the non-Maxwellian dusty plasma where lighter species behave Kappa and Cairns, all other assumptions described above will remain the same. Hence the number densities for Kappa distributed electrons and ions can be represented as given by Eq. (6) Poisson equation is given as

$$Z_d n_d=\frac{T_e}{4\pi e^2}\nabla^2\Phi-\left(\frac{\kappa-1/2}{\kappa-3/2}\right)N_0\Phi \tag{90}$$

where $N_0=\frac{n_{0i}T_e+n_{0e}T_i}{T_i}$, to obtain above equation, (90) has been expanded for $\kappa\geq 3$ ($e\phi<k_B T_{e,i}$), otherwise for $\kappa<3$, both higher and lower order terms would have been comparable so could not be ignored. Just like previous cases, we obtain

$$\left(\frac{\partial}{\partial t}+v_0\frac{\partial}{\partial z}+\frac{c}{B_0}\hat{\mathbf{z}}\times\nabla\phi\cdot\nabla\right)\left[\frac{T_e}{4\pi e^2}\nabla^2\Phi-\left(\frac{\kappa-1/2}{\kappa-3/2}\right)N_0\Phi\right]+\frac{Z_d cT_e}{eB_0}\hat{\mathbf{z}}\times\nabla\phi\cdot\nabla n_{d0}$$
$$+\frac{Z_d n_{d0}cT_e}{eB_0\Omega_d}\left(\frac{\partial}{\partial t}+v_0\frac{\partial}{\partial z}+\frac{cT_e}{eB_0}\hat{\mathbf{z}}\times\nabla\Phi\cdot\nabla\right)\nabla^2\Phi+Z_d n_{d0}\frac{\partial v_{dz}}{\partial z}=0 \tag{91}$$

And in normalized potential form (91) can be expressed as

$$\left(\frac{\partial}{\partial t} + v_0 \frac{\partial}{\partial z}\right) \mathbf{V}_{dz} + \frac{cT_e}{eB_0}\{\Phi, \mathbf{v}_{dz}\} + \frac{cT_e}{eB_0}\frac{1}{r}\frac{\partial \Phi}{\partial \theta}\frac{\partial v_0}{\partial r} = \frac{Z_d T_e}{m_d}\frac{\partial \Phi}{\partial z} \qquad (92)$$

Continuity equation can be written as:

$$\left[-\left(\frac{\kappa - 1/2}{\kappa - 3/2}\right)N_0 + \lambda_{DD}^2 \nabla_\perp^2 + \rho_d^2 \nabla^2\right]\frac{\partial \Phi}{\partial t} \times$$

$$-v_0\left(\frac{\kappa - 1/2}{\kappa - 3/2}\right)N_0\frac{\partial \Phi}{\partial z} + \Omega_d \rho_d^2(\lambda_D^2 + \rho_d^2)\{\Phi, \nabla^2 \Phi\} + \Omega_d \rho_d^2 \frac{1}{r}\frac{\partial}{\partial r}\ln n_{d0}\frac{\partial \Phi}{\partial \theta} + \frac{\partial V_{dz}}{\partial z} = 0 \quad (93)$$

or

$$\left[\frac{\left(\frac{\kappa-1/2}{\kappa-3/2}\right)N_0 - \frac{V_{*n}}{\Omega_0} + \frac{\alpha_0(2b\Omega_d \rho_d^2 - c_{sd}^2 \alpha_0)}{\Omega_0^2}}{(\lambda_{DD}^2 + \rho_d^2)}\right] L_\phi \Phi = L_\phi \nabla^2 \Phi \qquad (94)$$

where $L_\Phi = \frac{\partial}{\partial \eta} - HD_\Phi$ and above equations reduce to the well-known equation

$$\nabla_\perp^2 \Phi + \Lambda^2 \Phi = Cr^2 \qquad (95)$$

where $C = \frac{C_1}{2H'}$ and

$$C_1 = \frac{-\left(\frac{\kappa-1/2}{\kappa-3/2}\right)N_0 + \frac{V_{*n}}{\Omega_0} - \frac{\alpha_0(2b\Omega_d \rho_d^2 - c_{sd}^2 \alpha_0)}{\Omega_0^2}}{(\lambda_{DD}^2 + \rho_d^2)} - \Lambda^2 \qquad (96)$$

Solution is given as:

$$\Phi(r, \eta) = \Phi_m J_n\left(\frac{k_a r}{r_0}\right)\cos(n\eta) + \frac{A_0^{(\kappa)^2} r_0^2}{k_a^2}\left(r^2 - \frac{4r_0^2}{k_\alpha^2}\right) \qquad (97)$$

where

$$A_0^{(\kappa)} = \frac{1}{2H'}\left[\frac{-\left(\frac{\kappa-1/2}{\kappa-3/2}\right)N_0 + \frac{V_{*n}}{\Omega_0} - \frac{\alpha_0(2b\Omega_d \rho_d^2 - c_{sd}^2 \alpha_0)}{\Omega_0^2}}{(\lambda_{DD}^2 + \rho_d^2)} - \frac{k_a^2}{r_0^2}\right] \qquad (98)$$

The general solution of (97) represents that global vortex profile is usually determined by rotation frequency $\Omega_0$, radius $r_0$ and maximum amplitude $\Phi_m$.

Now for the case of Cairns distributed particles following expressions will be used. After following along the same steps given in previous cases, we finally obtain

$$\Phi(r, \eta) = \Phi_m J_n\left(\frac{k_a r}{r_0}\right)\cos(n\eta) + \frac{A_0^{(\Gamma)^2} r_0^2}{k_a^2}\left(r^2 - \frac{4r_0^2}{k_\alpha^2}\right) \qquad (99)$$

where

$$A_0^{(\Gamma)} = \frac{1}{2H}\left[\frac{-\left(1 - \frac{4\Gamma}{1+3\Gamma}\right)N_0 + \frac{V_{*n}}{\Omega_0} - \frac{\alpha_0(2b\Omega_d \rho_d^2 - c_{sd}^2 \alpha_0)}{\Omega_0^2}}{(\lambda_{DD}^2 + \rho_d^2)} - \frac{k_a^2}{r_0^2}\right] \qquad (100)$$

## X. QUANTITATIVE ANALYSIS AND CONCLUSIONS

For the pictorial view of the modon formation, we choose some typical dusty plasma parameters taken from Shukla and Mamum 2002[7] such as $m_i = 1.67 \times 10^{-27}$, $m_d = 1 \times 10^{-6} g$, $n_{do} = 10^{-7} cm^{-3}$, $n_{eo} = 10 cm^{-3}$, $n_{io} = 10 n_{eo}$, $B = 0.2G$, $T_e = 10^6 K$ and $T_i = 0.1 T_e$. These chosen parameters are also consistent with the assumption of cold dust and that the dust ion collision frequency is smaller than the dust gyro frequency.

As can be seen from the contour plots of Eq. (64) which is when dust is active presented in Figs. (1a, 1b, 1c & 1d) for $n = 1, 2, 3, 4$, respectively, motion and size of vortex is decided by the mode number $n$, which should be smaller than the cross section of plasma column.

For a mesospheric dusty plasma we have plotted shadow graphs of 3D view of normalized electrostatic potential $n_{d0} = 10^3$ cm$^{-3}$, $T_e \simeq T_i = 0.033$ eV, $Z_d = 100 - 500$, $\Omega_d = 0.1$, $T_s = 1.01 T_e$ Figs. (2a, 2b, 2c, 2d) show modons for n= 1, 2, 3 and 4 respectively.

We have also obtained modon formation for non-Maxwellian i.e., both Kappa and Cairns distributed plasma. Also we note how increase in the value of the dust charge (no of electrons) influence the coefficient $A_0$, as can be seen in the Fig. (3), increasing the charge number reduces the strength of the coefficient $A_0$ linearly. In Figs. (3 & 4), we also note how variations in the Kappa and Gamma indices impact the coefficient $A_0$ versus dust charge number ($Z_d$), respectively. As can be noted in the green line of Fig. (3), for the fixed value of $\kappa(=3)$, upon increasing the values of $Z_d$, strength of the $A_0$ is enhanced significantly. Also curves of coefficient reaches close to the Maxwellian (black line), when value of $\kappa$ increases. In Fig. (3) similar behaviors for the Cairns distributed ions and electrons has been observed, increase in the parameter $\Gamma(= 0.07, 0.3, 0.5)$, impacts strength of $A_0$ eventually heading towards Maxwellian curve for $\Gamma = 0.07$. Figs. (5a, 5b, 5c & 5d) for $n = 1, 2, 3, 4$, and fixed value of kappa $\kappa(= 5)$, demonstrates the contour plot of potential of the vortex in non-Maxwellain Kappa distributed plasma.

Superthermality is found to affect significantly the formation of the vortices. We report that for the ($\kappa = 2.8$ & 4), a significant difference has been observed on the vortex motion. Size of vortices decrease when superthermality is reduced and for the highly thermal plasma $\kappa = 2$, the vortex structure start disappearing.

However in case of Cairns distribution, when $\Gamma = .1, 0.2$, thermality increases vortices start disappearing. Turbulent or complex systems are usually composed of a large number of vortices at various scales, and the interaction among these vortices determines transport and evolution the behavior of the system.

In this paper, we have found a new mode, namely, the convective cell mode in the comet Halley plasma where role of ions is more significant and contribution of dust in the background plasma do not seem to affect it all. For this mode contribution of density and flow gradients is cancelled because of the presence of negative ions. For the linear analysis of this mode we obtain an eigen value equation whereas nonlinear analysis admits usual modons. The present analysis can be useful for interpreting certain astrophysical situations where vortex structures can be formed and as well as for the lab experiments on dusty plasma including fusion devices in particular for the edge turbulence and anomalous transport. The analysis of non-Maxwellain plasma is also useful in both lab and space plasma because velocity-space diffusion obeying power law distribution such as $(\epsilon/\epsilon_0)^{-\kappa}$ is an inherent feature of any lab plasma experiment and of crucial significance in the transport properties in fact superthermal lab plasma behavior has been observed and reported.

**Author's Contributions**

All authors contributed equally to this manuscript.

**Availability of Data**

The data that support the findings of this study are available from the corresponding author upon reasonable request.

**Figure Caption**s

Fig. 1(a, b, c & d): Profile of modons in the maxwellian dusty plasma for $n = 1, 2, 3, 4$.

Fig. (2): Shadow graph of 3D view of modons formation in the mesospheric dusty plasma.

Fig. (3): Coefficient $A_0$ versus $Z_d$ for a different values of spectral indices $\kappa$ ($= 2, 3, 5, 10$).

Fig. (4): Coefficient $B_0$ versus $Z_d$ for a different values of non-thermal particles i.e., $\Gamma = (0.07, 0.3, 0.5)$.

Fig. 5(a, b, c): Profile of vortices in the Maxwellian and non-maxwellain (fixed, $\Gamma(= 0.01)$, $\kappa(= 7)$) dusty plasma for $n = 1$.

Fig. (4d): A comparison of vortex formation for three different distribution functions.

## XI. APPENDEX

$$\hat{\mathbf{z}} \times \nabla \Phi \cdot \nabla \mathbf{V}_{dz} = \begin{vmatrix} 0 & 0 & 1 \\ \frac{\partial \Phi}{\partial r} & \frac{1}{r}\frac{\partial \Phi}{\partial \theta} & 0 \\ \frac{\partial \mathbf{V}_{dz}}{\partial r} & \frac{1}{r}\frac{\partial \mathbf{V}_{dz}}{\partial \theta} & \frac{\partial \mathbf{V}_{dz}}{\partial z} \end{vmatrix} = \{\Phi, \mathbf{V}_{dz}\}$$

where

$$\{\Phi, \mathbf{V}_{dz}\} = \frac{1}{r}\left[\frac{\partial \Phi}{\partial r}\frac{\partial \mathbf{V}_{dz}}{\partial r} - \frac{\partial \Phi}{\partial \theta}\frac{\partial \mathbf{V}_{dz}}{\partial r}\right]$$

is the Poisson bracket

$$\hat{\mathbf{z}} \times \nabla \Phi \cdot \nabla \left(\nabla^2 \Phi\right) = \begin{vmatrix} 0 & 0 & 1 \\ \frac{\partial \Phi}{\partial r} & \frac{1}{r}\frac{\partial \Phi}{\partial \theta} & 0 \\ \frac{\partial \nabla^2 \Phi}{\partial r} & \frac{1}{r}\frac{\partial \nabla^2 \Phi}{\partial \theta} & 0 \end{vmatrix} = \{\Phi, \nabla^2 \Phi\}$$

where

$$\{\Phi, \nabla^2 \Phi\} = \frac{1}{r}\left[\frac{\partial \Phi}{\partial r}\frac{\partial \nabla^2 \Phi}{\partial r} - \frac{\partial \Phi}{\partial \theta}\frac{\partial \nabla^2 \Phi}{\partial r}\right]$$

is the Poisson bracket

$$\hat{\mathbf{z}} \times \nabla \Phi \cdot \frac{\nabla n_{d0}}{n_{d0}} = \frac{1}{r}\frac{\partial}{\partial r}\ln n_{d0}\frac{\partial \Phi}{\partial \theta}$$

$$\hat{\mathbf{z}} \times \nabla v_0 \cdot \nabla \Phi = \frac{1}{r}\frac{\partial v_0}{\partial r}\frac{\partial \Phi}{\partial \theta}$$

$$\hat{\mathbf{z}} \times \nabla \mathbf{V}_{dz} \cdot \nabla \Phi = \frac{\partial \mathbf{V}_{dz}}{\partial r}\frac{1}{r}\frac{\partial \Phi}{\partial \theta}$$

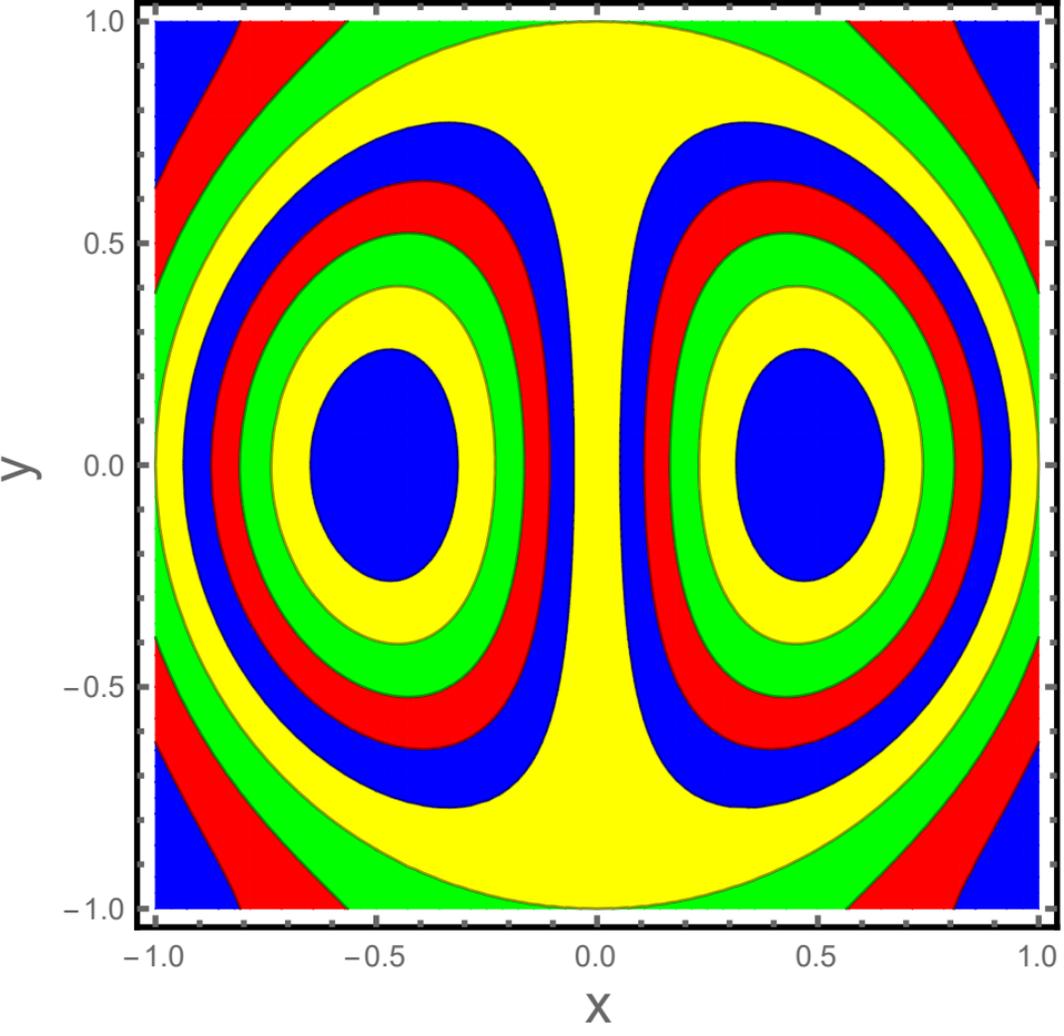

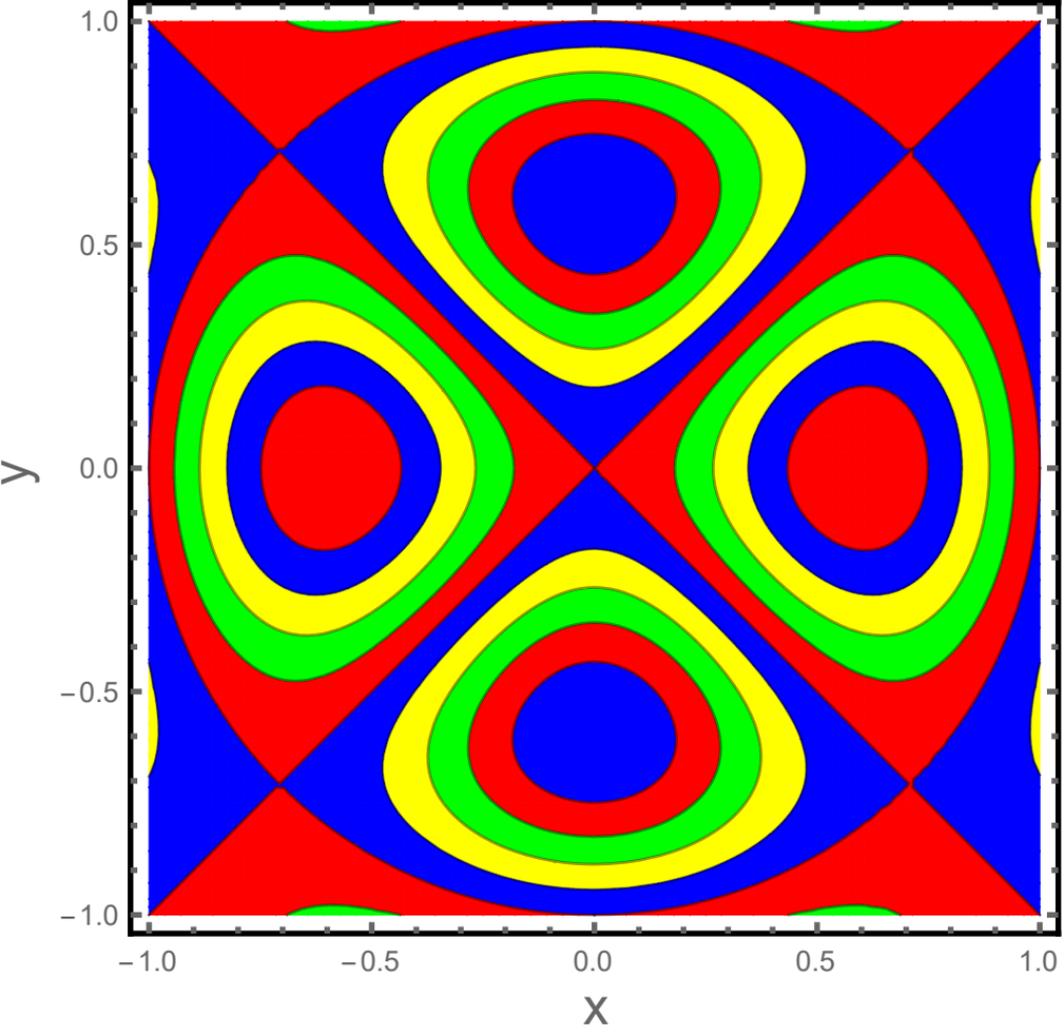

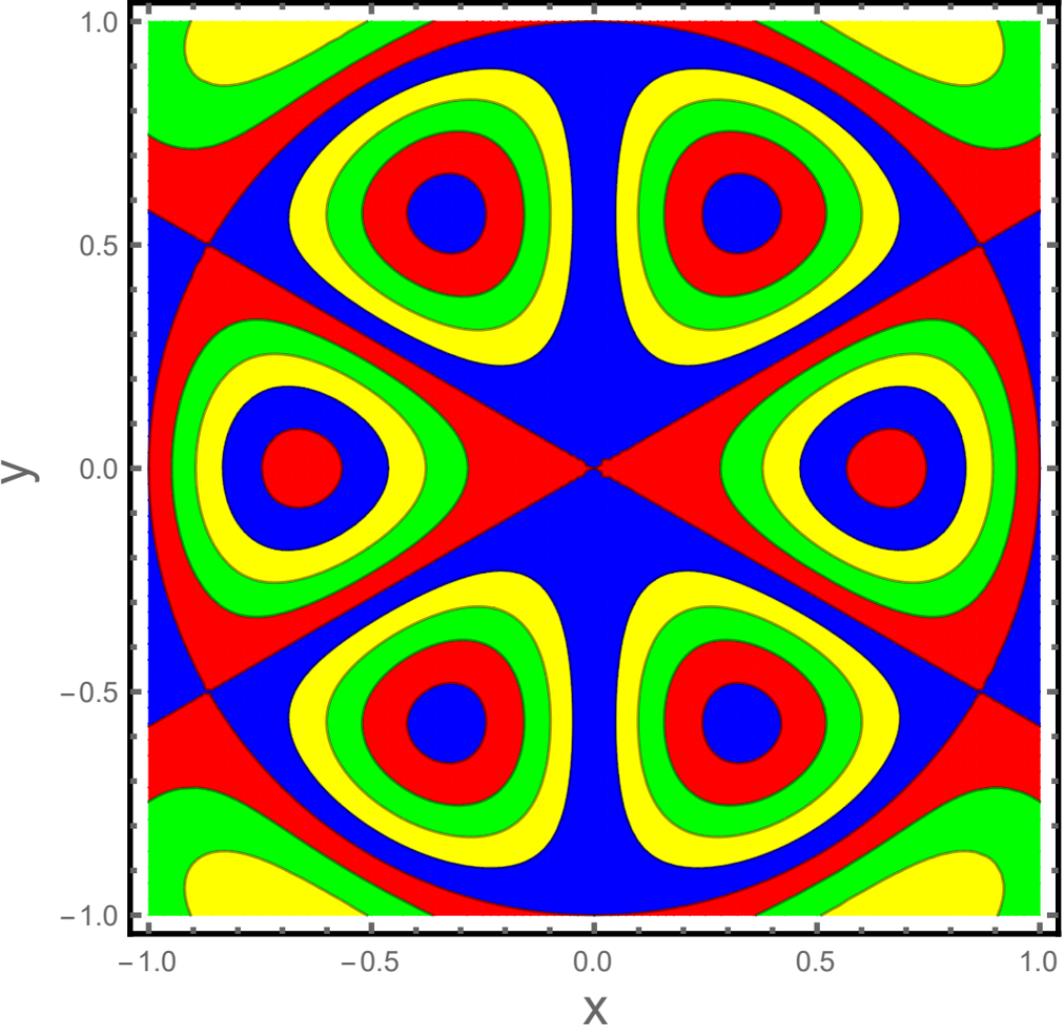

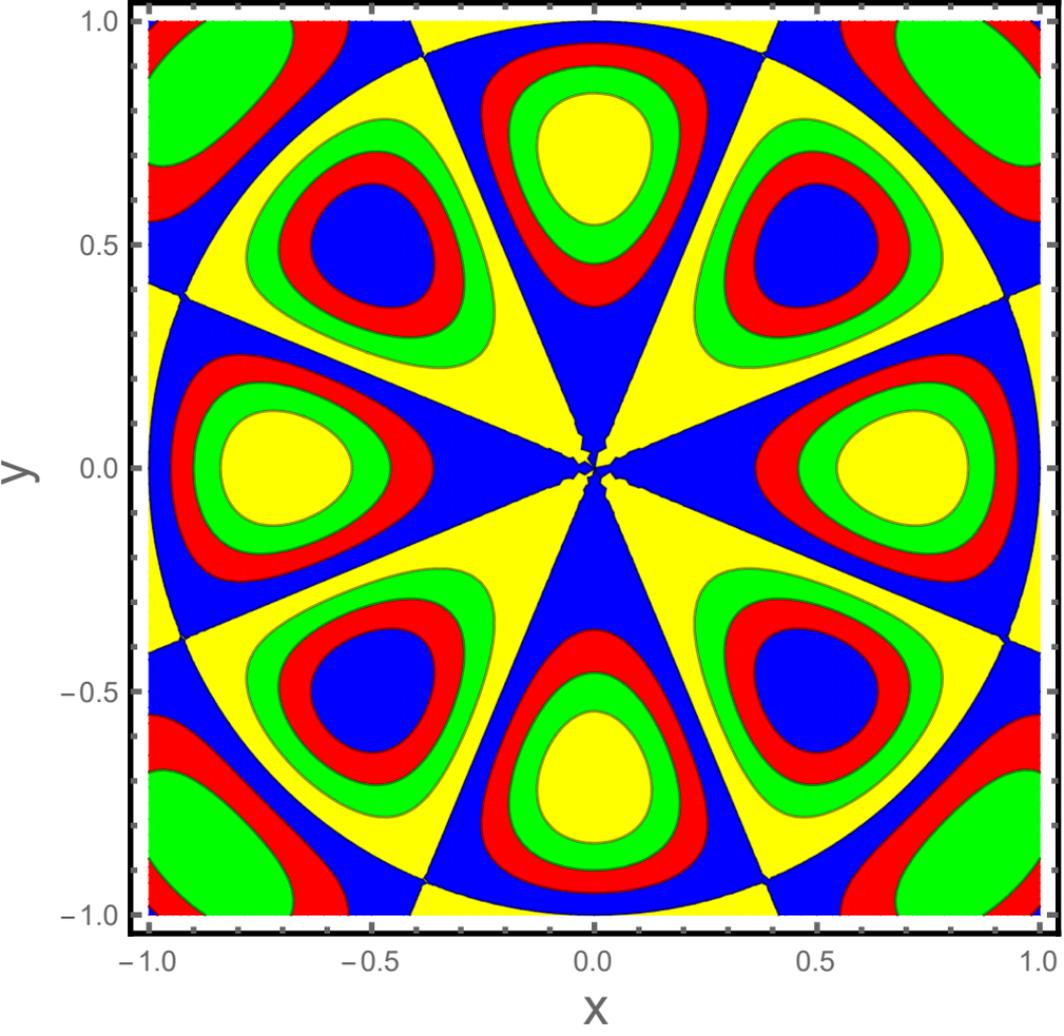

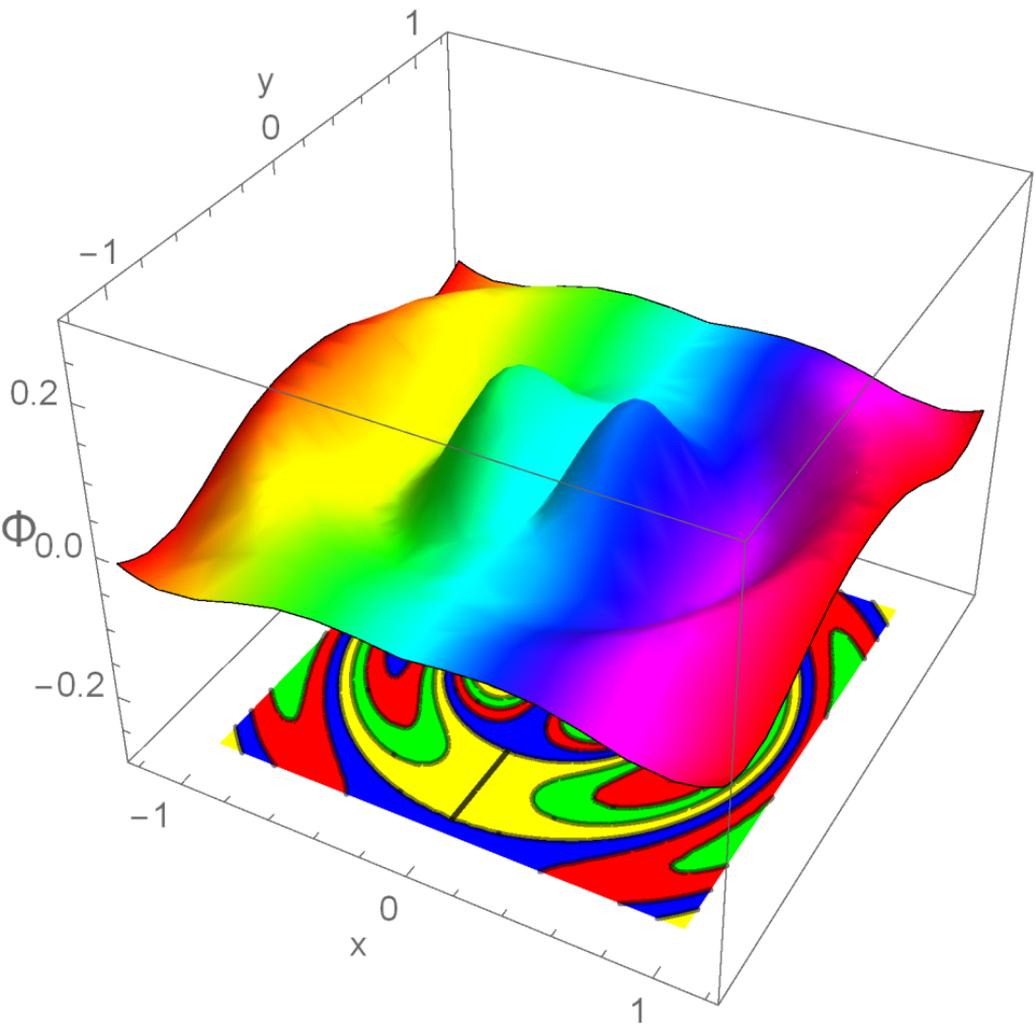

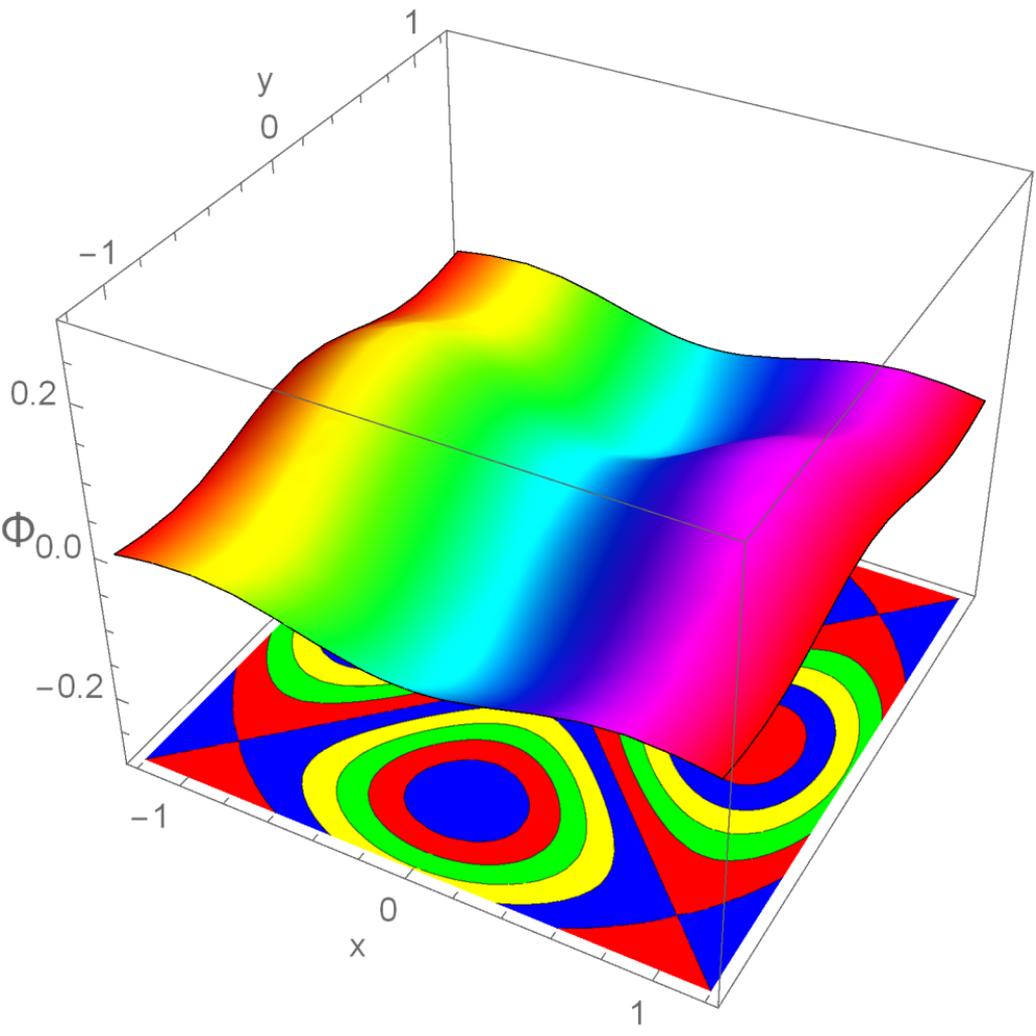

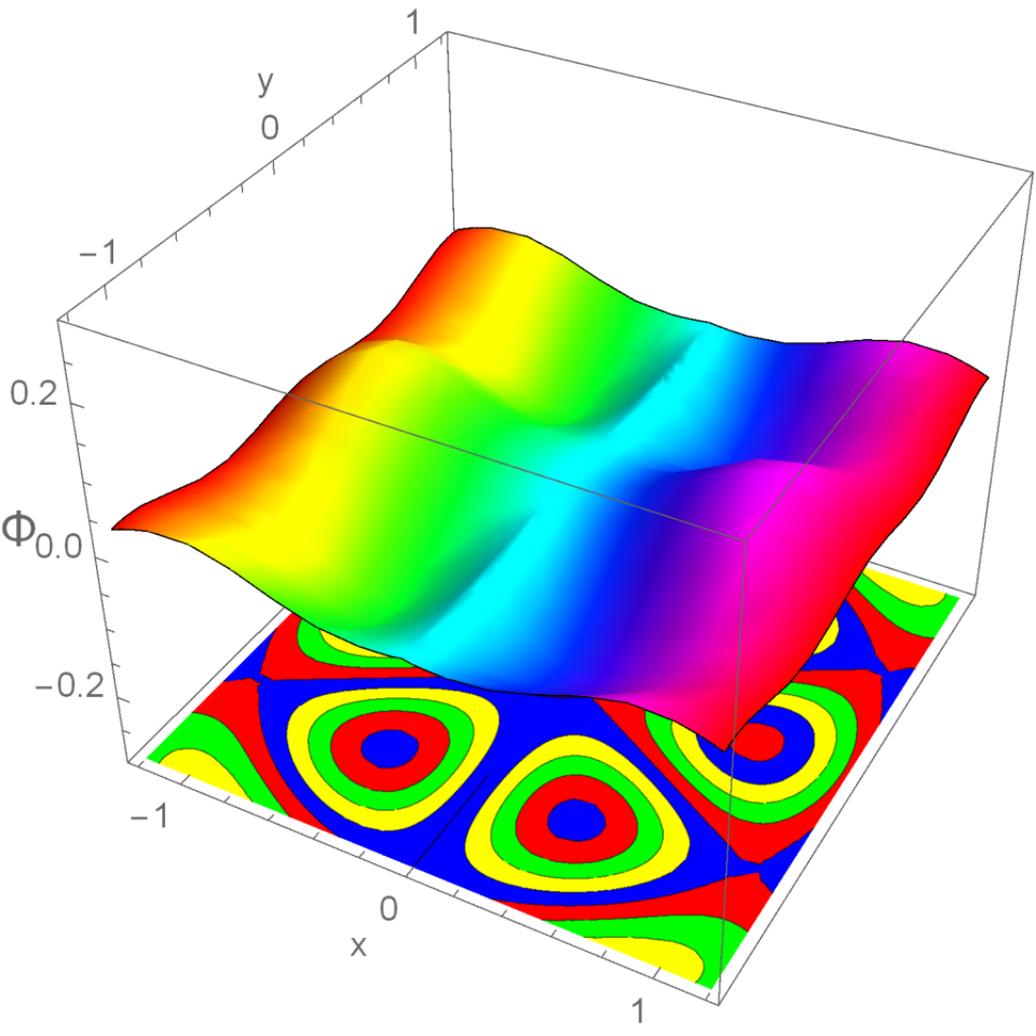

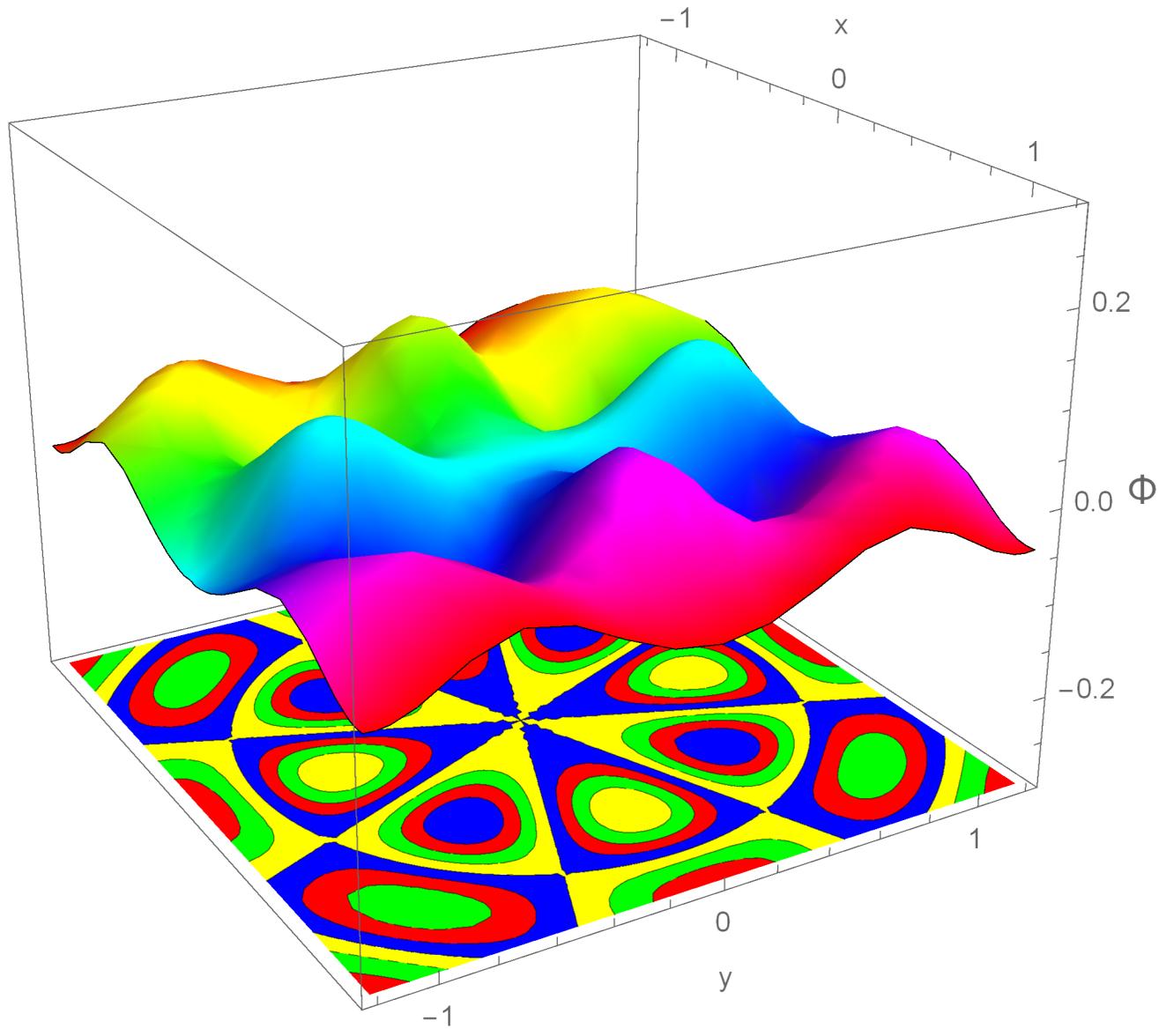

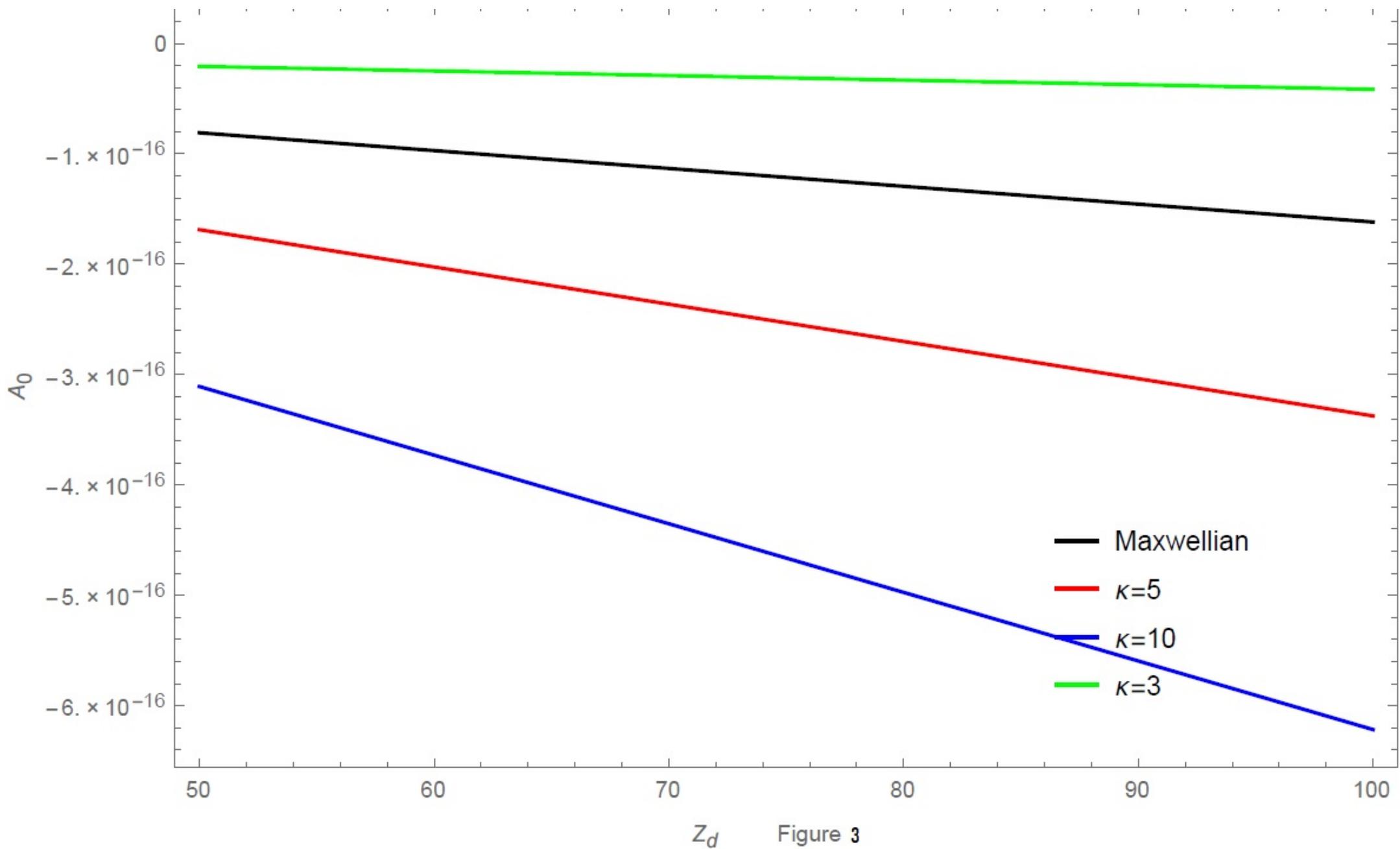

Figure 3

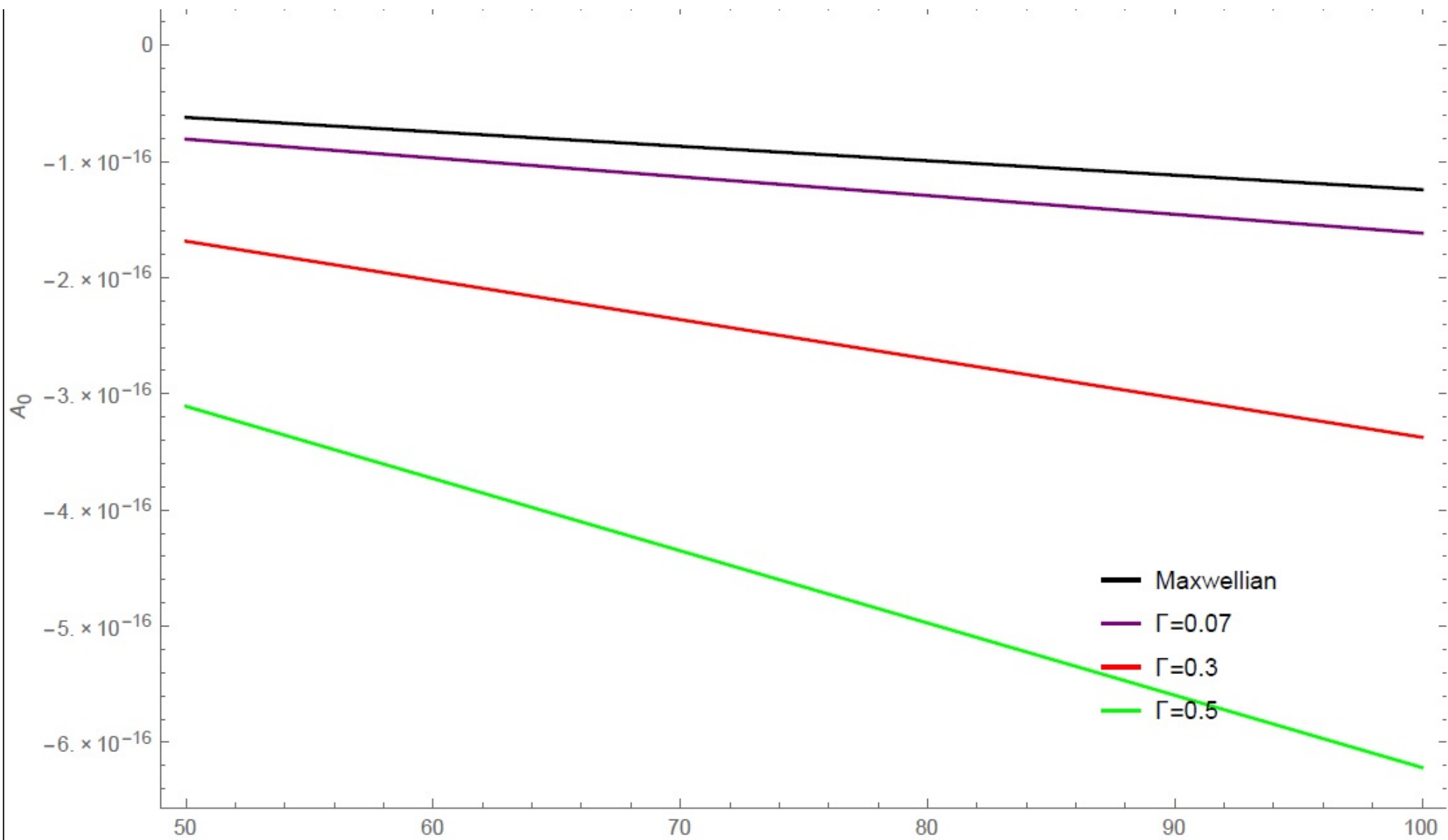
Figure 4

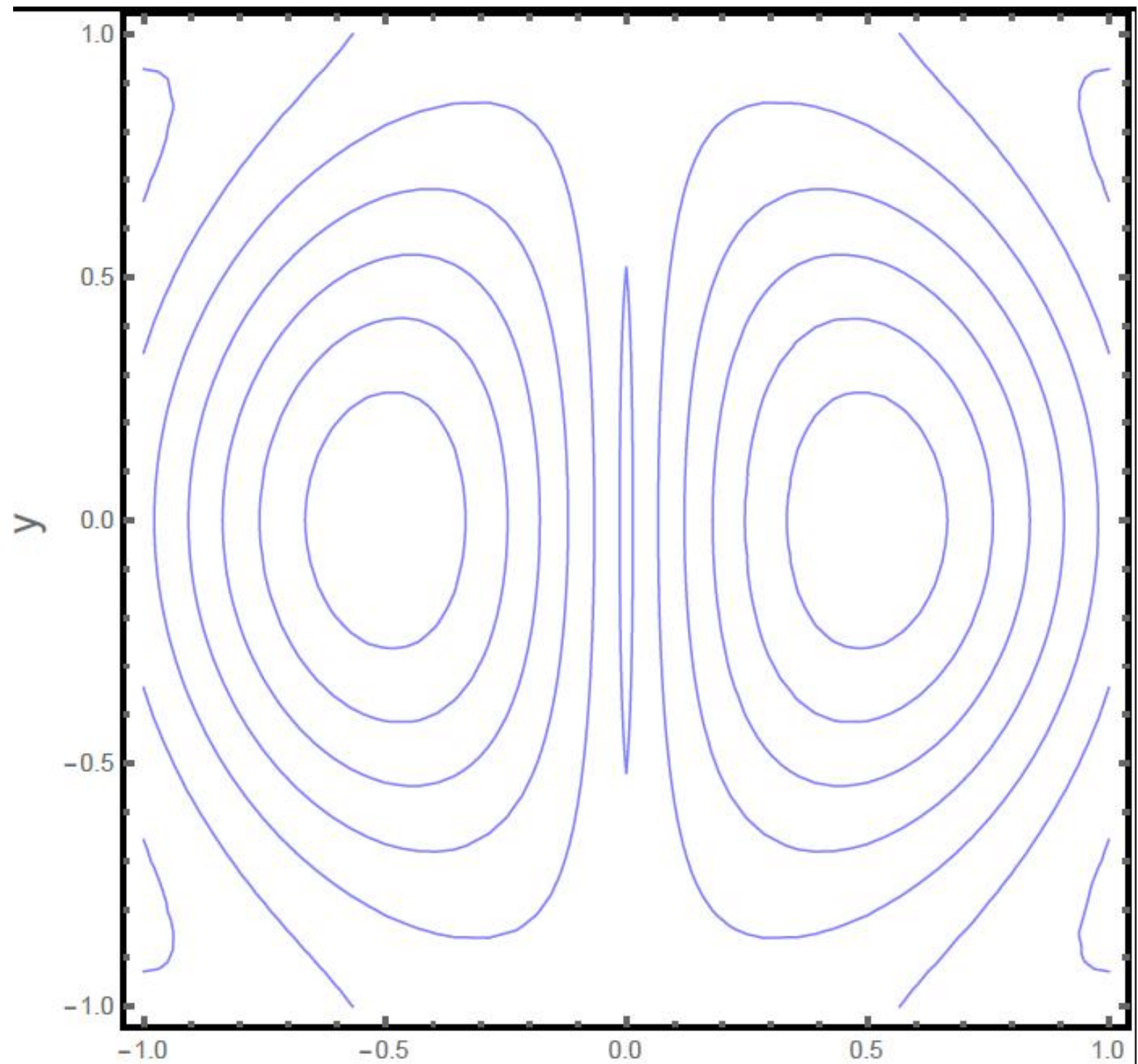

Fig. (5a) Maxwellain

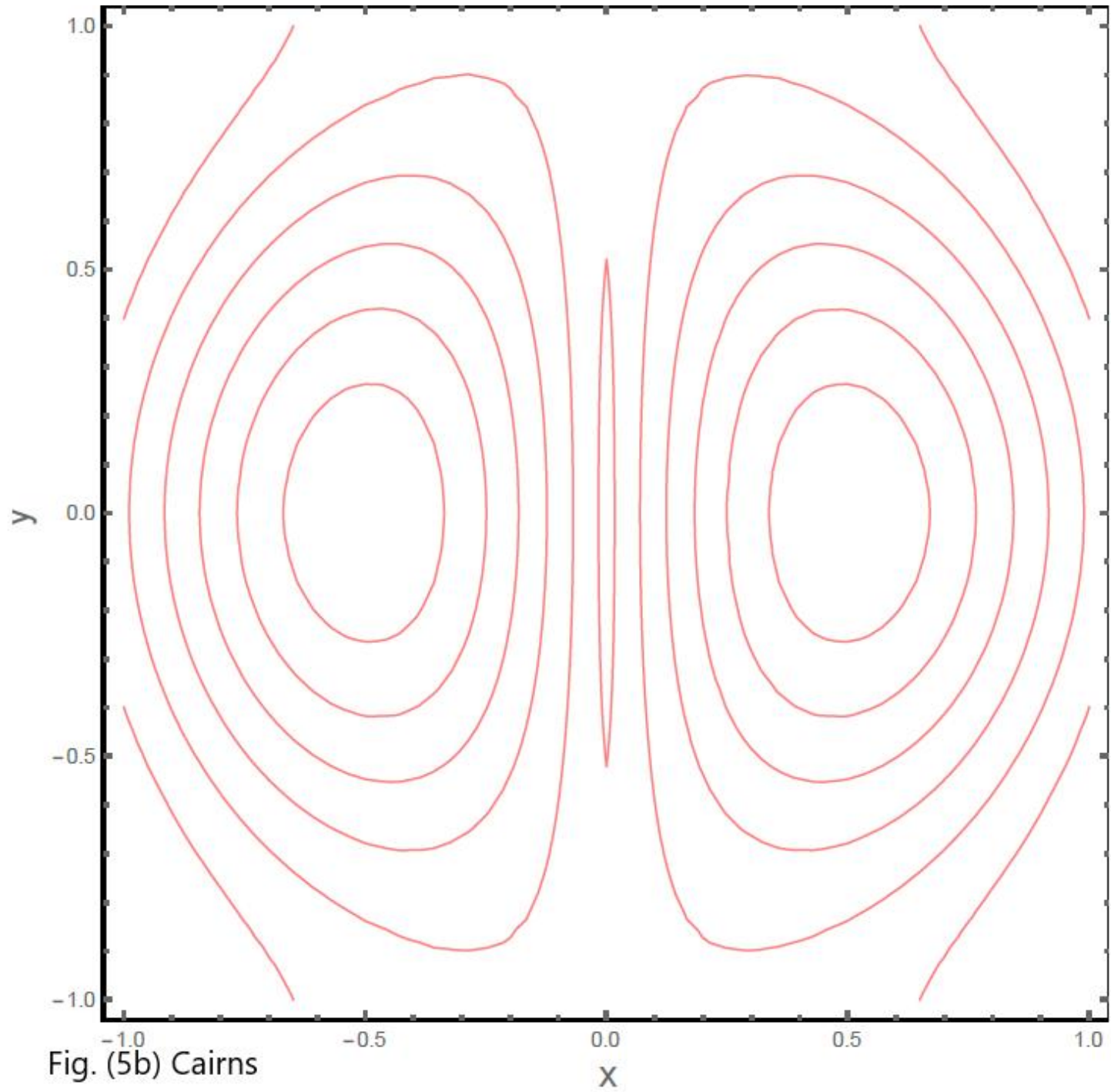

Fig. (5b) Cairns

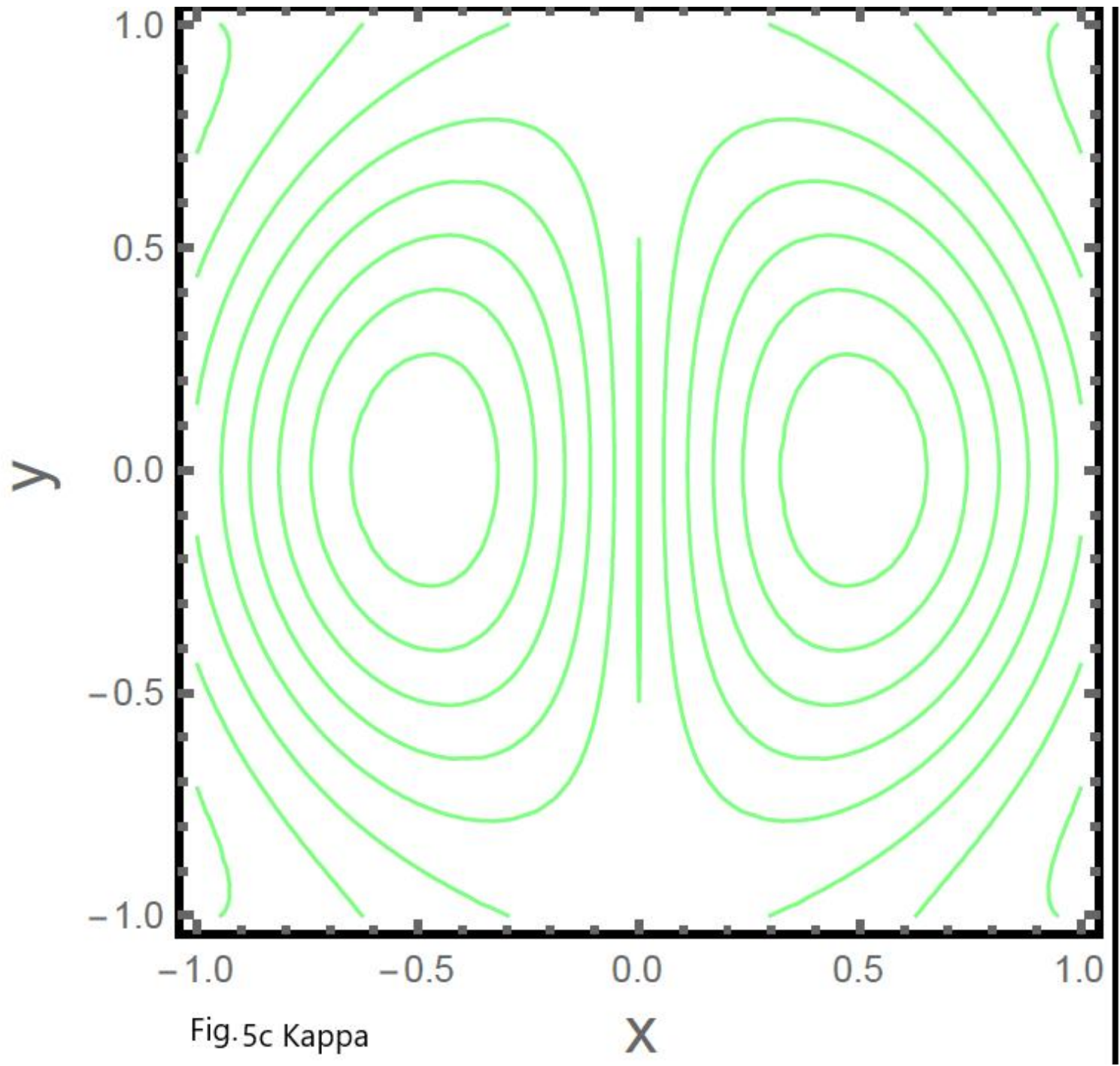
Fig. 5c Kappa

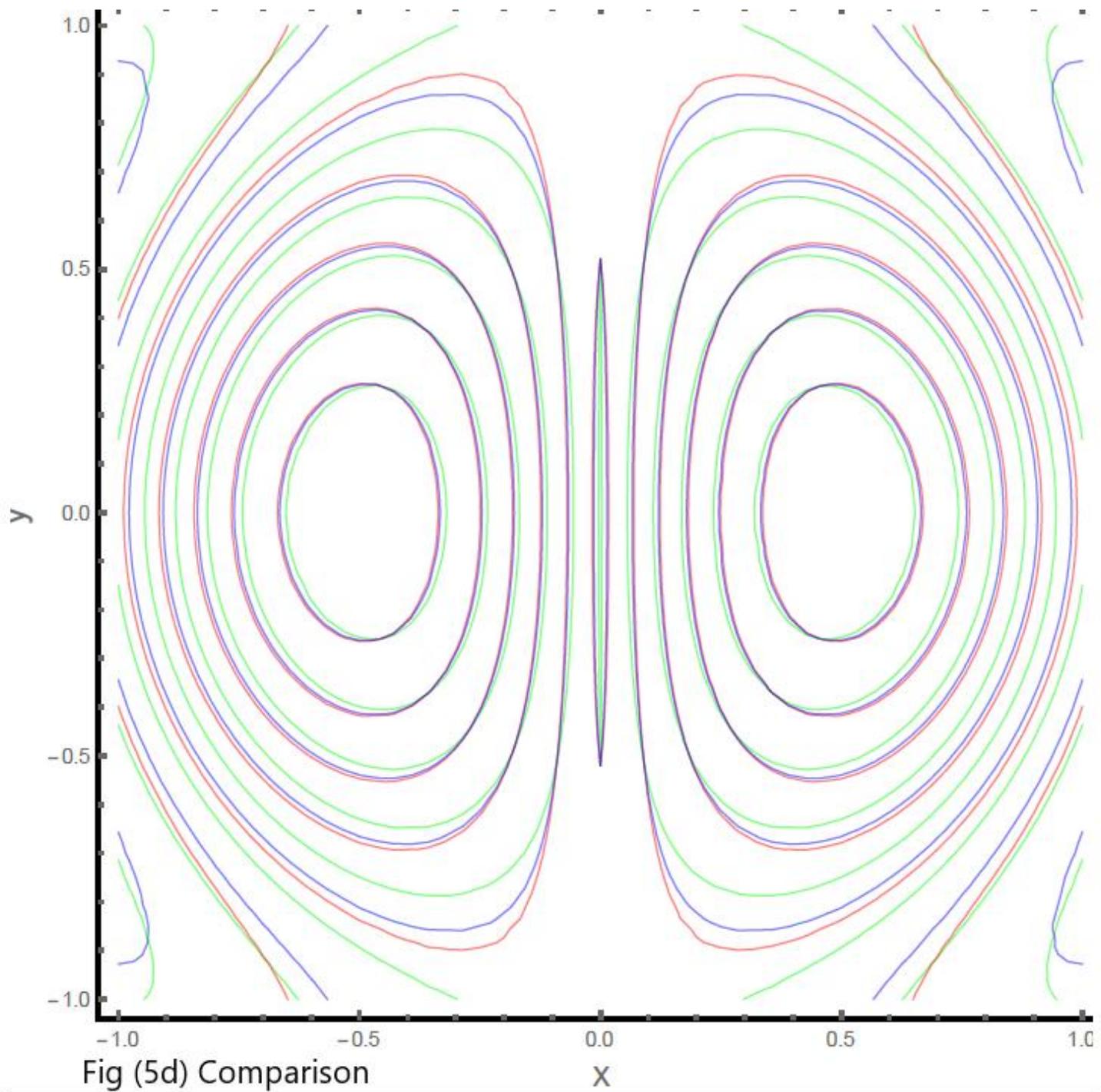

Fig (5d) Comparison